\documentclass[sigconf, nonacm]{acmart}

\usepackage[vlined,ruled,linesnumbered]{algorithm2e}
\usepackage{algorithmicx}
\usepackage{enumitem}

\usepackage{amsthm}
\theoremstyle{definition}
\newtheorem{definition}{Definition}
\newtheorem{example}{Example}[section]

\usepackage{multirow}

\newcommand\vldbdoi{XX.XX/XXX.XX}
\newcommand\vldbpages{XXX-XXX}
\newcommand\vldbvolume{14}
\newcommand\vldbissue{1}
\newcommand\vldbyear{2020}
\newcommand\vldbauthors{\authors}
\newcommand\vldbtitle{\shorttitle} 
\newcommand\vldbavailabilityurl{URL_TO_YOUR_ARTIFACTS}
\newcommand\vldbpagestyle{plain}

\newcommand{\htitle}[1]{\vspace{1mm} \noindent \textbf{#1}}
\begin{document}
\title{Cardinality Estimation for High Dimensional Similarity Queries with Adaptive Bucket Probing}

\author{Zhonghan Chen}
\affiliation{%
  \institution{The Hong Kong University of Science and Technology}
  \streetaddress{Clear Water Bay}
  \city{Hong Kong SAR}
  \state{China}
}
\email{zchenhj@connect.ust.hk}

\author{Qintian Guo}
\affiliation{%
  \institution{The Hong Kong University of Science and Technology}
  \streetaddress{Clear Water Bay}
  \city{Hong Kong SAR}
  \state{China}
}
\email{qtguo@ust.hk}

\author{Ruiyuan Zhang}
\affiliation{%
  \institution{Hong Kong Generative AI Research and Development Center (HKGAI)}
  \streetaddress{Clear Water Bay}
  \city{Hong Kong SAR}
  \state{China}
}
\email{zry@hkgai.org}

\author{Xiaofang Zhou}
\affiliation{%
  \institution{The Hong Kong University of Science and Technology}
  \streetaddress{Clear Water Bay}
  \city{Hong Kong SAR}
  \state{China}
}
\email{zxf@ust.hk}

\begin{abstract}
In this work, we address the problem of cardinality estimation for similarity search in high-dimensional spaces. Our goal is to design a framework that is lightweight, easy to construct, and capable of providing accurate estimates with satisfying online efficiency.
We leverage locality-sensitive hashing (LSH) to partition the vector space while preserving distance proximity. Building on this, we adopt the principles of classical multi-probe LSH to adaptively explore neighboring buckets, accounting for distance thresholds of varying magnitudes. To improve online efficiency, we employ progressive sampling to reduce the number of distance computations and utilize asymmetric distance computation in product quantization to accelerate distance calculations in high-dimensional spaces. In addition to handling static datasets, our framework includes updating algorithm designed to efficiently support large-scale dynamic scenarios of data updates.
Experiments demonstrate that our methods can accurately estimate the cardinality of similarity queries, yielding satisfying efficiency.
\end{abstract}

\maketitle

\pagestyle{\vldbpagestyle}
\begingroup\small\noindent\raggedright\textbf{PVLDB Reference Format:}\\
\vldbauthors. \vldbtitle. PVLDB, \vldbvolume(\vldbissue): \vldbpages, \vldbyear.\\
\href{https://doi.org/\vldbdoi}{doi:\vldbdoi}
\endgroup
\begingroup
\renewcommand\thefootnote{}\footnote{\noindent
This work is licensed under the Creative Commons BY-NC-ND 4.0 International License. Visit \url{https://creativecommons.org/licenses/by-nc-nd/4.0/} to view a copy of this license. For any use beyond those covered by this license, obtain permission by emailing \href{mailto:info@vldb.org}{info@vldb.org}. Copyright is held by the owner/author(s). Publication rights licensed to the VLDB Endowment. \\
\raggedright Proceedings of the VLDB Endowment, Vol. \vldbvolume, No. \vldbissue\ %
ISSN 2150-8097. \\
\href{https://doi.org/\vldbdoi}{doi:\vldbdoi} \\
}\addtocounter{footnote}{-1}\endgroup

\ifdefempty{\vldbavailabilityurl}{}{
\vspace{.3cm}
\begingroup\small\noindent\raggedright\textbf{PVLDB Artifact Availability:}\\
The source code, data, and/or other artifacts have been made available at \url{https://github.com/OscarC9912/simQ_hd_card_estimator}.
\endgroup
}

\section{Introduction}
\label{sec: introduction}
In relational database systems\cite{postgresql}, cardinality estimation\cite{DBLP:journals/pvldb/KieferHBM17, DBLP:journals/pvldb/ShetiyaTK020, DBLP:conf/sigmod/GetoorTK01, DBLP:journals/vldb/GunopulosKTD05, DBLP:conf/sigmod/DeshpandeGR01, DBLP:journals/pvldb/YangLKWDCAHKS19, DBLP:journals/pvldb/YangKLLDCS20, DBLP:journals/pvldb/HanWWZYTZCQPQZL21} is an important component of query optimization\cite{10.1145/234313.234367,DBLP:conf/pods/Chaudhuri98} that has been studied for decades, which provides a fast estimation towards the number of output rows of the query, in order to select the query execution plan of the smallest latency. Recently, with the prevalence of vector database\cite{milvus, pgvector, pase, weaviate, chroma, qdrant, lancedb, vearch, pinecone, AnalyticDB-V, vdb_survey} and semantic operators\cite{patel2024semanticoperators, DBLP:journals/pacmmod/BalakaAWGKF25}, cardinality estimation for similarity query in the high dimensional space (CE4HD) \cite{10.14778/3712221.3712224, DBLP:journals/pvldb/QinX18, DBLP:conf/edbt/MattigFBS18, DBLP:conf/icde/QinWXWLI18, DBLP:conf/sigmod/KimJSHCC22, DBLP:conf/sigmod/WangXQ0O00I21, DBLP:conf/sigmod/WangXQ0SWO20, DBLP:conf/icml/WuCN18, DBLP:conf/sigmod/Sun0021}, has become popular in database research, where we aim to efficiently estimate the number of points $x$ that are within distance threshold $\tau \in \mathbb{R}$ from query point $q\in \mathbb{R}^d$, given a range query $(x\in \mathbb{R}^d, \tau \in \mathbb{R})$. In terms of application, cardinality estimator of similarity query can be used to optimize the execution plan of semantic operators that interacts with large language models\cite{LLaMA,llama2,GPT4,deepseek,Gemini}, where we can efficiently estimate the number of interactions with LLM without actual execution.


Recent works on CE4HD\cite{DBLP:conf/sigmod/Sun0021, DBLP:conf/sigmod/WangXQ0SWO20, DBLP:conf/sigmod/WangXQ0O00I21, 10.14778/3712221.3712224} primarily utilize deep neural networks (DNN) to predict cardinality given similarity queries. Technically, these approach generates a feature representation derived from the query vector, the distance threshold, and various dataset properties. This representation is then used to train a learned model, which in turn is employed to estimate the cardinality of vector similarity queries. However, learned approach suffers from several key disadvantages in practice. Firstly, the performance of DNN-based predictors relies heavily on the quality of training data, whose performance degrades dramatically if training data is not properly tailored. Secondly, the DNN-based estimators require the offline construction that takes substantial amount of time. 
For example, SimCard\cite{DBLP:conf/sigmod/Sun0021} partitions the dataset into smaller clusters (e.g. $100$-$200$ clusters), where the training label is computed with respect to each cluster, and an estimation model will be trained for each of them, in order to derive the estimation for the current local area. For this pipeline, even with performant GPUs, such process is still time-consuming and requires many computational resources. Thirdly, DNN-based estimators suffer from a destitute of explainability, and exhibits instability, demonstrating large performance discrepancies across different datasets.
Motivated by the aforementioned disadvantages of using deep neural networks or learning-based frameworks, we aim to design a lightweight solution that can perform well in both static and dynamic scenarios, with theoretical guarantees to support its robustness and effectiveness.

Generally speaking, our primary design guideline is that the estimator should be light-weighted and requires little computational resources, so that the framework will be efficient for offline construction, online estimation, as well as large-scale dynamic data updates. To achieve these objectives, our framework is built upon the locality sensitive hashing (LSH)\cite{e2lsh}, which is a light-weighted vector index for approximate nearest neighbor (ANN) search, with extra optimization for improving efficiency with product quantization and progressive sampling. 

\begin{figure*}[t]
    \centering
    \includegraphics[width=\linewidth]{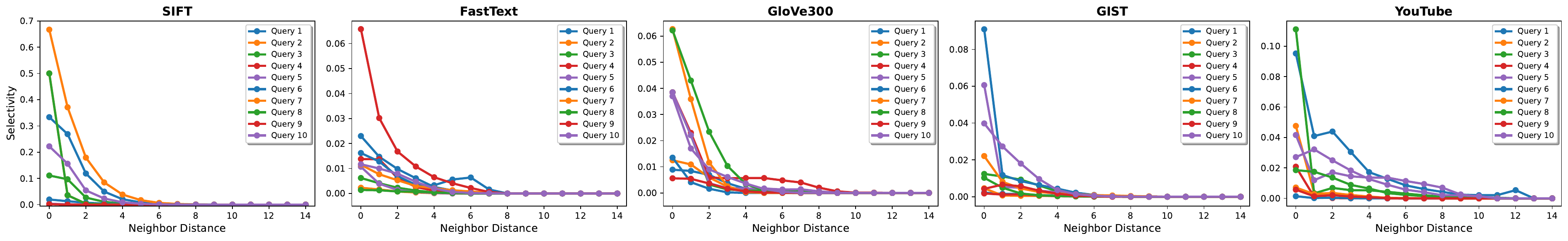}
    \caption{Motivation of our work: $x$-axis is the distance between the central bucket $\mathcal{B}_{central}$ with its $k$-step neighbor $\mathcal{N}_k$, where $k\in[0,14]$ and $k$ is the hamming distance, and $y$-axis is the selectivity in $\mathcal{N}_k$. We use $5$ datasets, each containing $10$ queries, for demonstration. As $\mathcal{N}_k$ becomes more distant from $\mathcal{B}_{central}$, the selectivity of the neighbor decreases, which means that closer neighbor are more likely to contain points that satisfies the similarity query in context of cardinality estimation.}
    \label{fig:motivation_dist_sel}
\end{figure*}

In ANN search, the multi-probe LSH\cite{DBLP:conf/vldb/LvJWCL07} claims that: \textit{given a query that is hashed into a particular bucket containing its nearest neighbor, it is highly probable that adjacent buckets also contain the nearest neighbor of the query}, which is the result of hard boundary problem in LSH. In cardinality estimation of similarity query in high dimensional space, we observe a similar tendency, which is demonstrated in Figure \ref{fig:motivation_dist_sel}. The overall intuition and motivation illustrated in Figure \ref{fig:motivation_dist_sel} is that as the neighboring buckets become more distant, the possibility for a neighbor to contain the points within the distance threshold decreases. The intuition here will be formalized and discussed in Section \ref{subsec: Neighboring-based Probing Strategy.}. Motivated by this observation, we propose a \textit{neighboring-based adaptive bucket probing strategy} for efficient cardinality estimation for similarity queries.

Technically, given a query $\mathcal{Q} =(x\in \mathbb{R}^d, \tau\in\mathbb{R})$, we firstly use LSH functions to find the central hash bucket $\mathcal{B}_{central}$ of the query, and then we adaptively probe the $\mathcal{N}_{k}$, which is the $k$-th step neighbor of $\mathcal{B}_{central}$, where $\mathcal{N}_{k}$ consists of all hash buckets whose hash code is at $k$ distance away in hamming space. In addition, the maximum value of $k$ is determined by the number of hash functions in the LSH index. In cardinality estimation for high dimensional similarity queries in high dimensional space, the distance computation is the bottleneck for efficiency. We further mitigate the bottleneck from two perspectives: 1). employing progressive sampling to reduce the number of distance computations, and 2). leveraging asymmetric distance computation in product quantization to improve the efficiency of distance calculation. 

A primary characteristic of cardinality estimation is that queries are associated with distance thresholds of widely varying magnitudes: some thresholds may yield only a few results, while others can return thousands of qualifying points. In the bucket probing scheme with LSH index, the magnitudes of distance threshold determines if distant neighbors should be probed or estimating within $\mathcal{B}_{central}$ is sufficient. However, the challenge is that $\tau$ is unknown before query arrives, which means we need to efficiently understand the magnitude of $\tau$ online dynamically. To tackle the challenge, we design a selectivity-based early stopping strategy to make our algorithm \textit{adaptive} to distance thresholds at various magnitudes, where the early stopping conditions are bounded with theoretical guarantee. The major contributions of paper are as follows:\vspace{-1mm}

\begin{itemize}[left=0pt]
    \item Design: we propose a locality-sensitive hashing (E$2$LSH) based framework to answer the cardinality estimation problem in high dimensional Euclidean space with problem-specific optimization.
    \item Adaptive Probing: we design an efficient \textit{neighboring-based adaptive bucket probing strategy} to dynamically probe the hash buckets and adjust the number of buckets to explore given distance thresholds at different magnitudes, according to a selectivity based early termination condition with theoretical guarantee. 
    \item Optimization: we integrate problem-specific optimization with two aspects. We develop an \textit{adaptive progressive sampling strategy} to reduce the number of distance computation needed and apply the \textit{asymmetric distance in product quantization (PQ)} to improve the efficiency of distance computation.
    \item Data Updates: we further leverage the advantage of our framework to support large-scale data updates, where we provide detailed explanation and algorithms for updating each of the component of the framework and experimental evaluations. 
\end{itemize}

The paper is organized as follows. Section \ref{sec: prelimiary} formally defines the problem and introduces the necessary preliminaries of our framework. Section \ref{sec: related work} reviews recent advances in cardinality estimation for similarity search. Section \ref{sec: framework} presents the core ideas of our framework along with tailored pseudocode for each component of the proposed algorithms. Section \ref{sec: dynamic data} extends the framework to handle dynamic data updates. Section \ref{sec: evaluation} provides an extensive experimental evaluation of the algorithms. Finally, Section \ref{sec: conclusion} concludes the paper and outlines three promising directions for future research.

\section{Preliminaries}
\label{sec: prelimiary}
\begin{table}
\centering
\label{tab:notations}
\scalebox{1.1}{
	\begin{tabular}{|c|c|}
		\hline
		\textbf{Notation} & \textbf{Description}\\ 
		\hline 
		$N$ &  The cardinality / size of dataset\\ \hline 
            $d$ & The dimensionality of dataset\\ \hline 
            $\mathcal{D} \in \mathbb{R}^{N\times d}$ & The dataset\\ \hline 
            $x, q \in \mathbb{R}^d$ & query / data point in $d$-dimensional space\\ \hline 
            $\tau \in \mathbb{R}$ & distance threshold in vector range search\\ \hline 
            $dist_{s}$ & distance function in $s$ space\\ \hline 
             $\mathcal{B}_{central}$ & hash bucket that a point being mapped\\ \hline 
             $\mathcal{N}_k$ & the $k$-Step Neighbor to $\mathcal{B}_{central}$\\ \hline 
             $\mathcal{I}$ & Locality-Sensitive Hashing Index\\ \hline 
             $\mathcal{Q}$ & Product Quantization Index\\ \hline 
             $\mathcal{C}$ & An array storing all hash code of hash table\\ \hline
             $\mathcal{P}$ & Neighbor Lookup Table \\ 
		\hline 
    \end{tabular}}
    \caption{List of Key Notations.}
\end{table}
\subsection{Problem Definition}



Vector similarity search is a fundamental operation in a wide range of machine learning and data analysis applications, such as information retrieval, recommendation systems, and computer vision. 
\begin{definition}[\textbf{Similarity Search}]
    Given a vector dataset $\mathcal{D} \in \mathbb{R}^{N\times d}$, a range query $(q, \tau)$, where $q \in \mathbb{R}^d$ and $\tau \in \mathbb{R}$, and a distance function $dist$, the vector similarity search returns all the $x \in \mathcal{D}$, whose $dist(q, x) \leq \tau$; that is $\{x | dist(x, q) \leq \tau, x\in \mathcal{D}\}$.
\end{definition}

This formulation captures a broad class of similarity search problems under various distance metrics (e.g., Euclidean, cosine, Manhattan). It is often referred to as a range query in geometric or metric space indexing.


\begin{definition}[\textbf{Cardinality Estimation for Similarity Search}]
    Given a vector dataset $\mathcal{D} \in \mathbb{R}^{N\times d}$, a range query $(q, \tau)$, where $q \in \mathbb{R}^d$ and $\tau \in \mathbb{R}$, and a distance function $dist(\cdot, \cdot)$, cardinality estimation for similarity search gives the number of data point in $\mathcal{D}$ whose distance to $q$ are no greater than $\tau$, i.e., $|\{x | dist(x, q) \leq \tau, x\in \mathcal{D}\}|$.
\end{definition}

{In the research of cardinality estimation for similarity search, previous works emphasized on one\cite{DBLP:conf/icde/QinWXWLI18, DBLP:conf/icml/WuCN18, DBLP:conf/edbt/MattigFBS18} or multiple\cite{DBLP:conf/sigmod/Sun0021, 10.14778/3712221.3712224, DBLP:conf/sigmod/WangXQ0O00I21, DBLP:conf/sigmod/WangXQ0SWO20} distance functions, such as hamming distance, angular distance, edit distance, Jaccard distance, as well as Euclidean distance. In our work, we focus on the estimation in Euclidean space.}

\begin{definition}[\textbf{Euclidean Distance}]
    The Euclidean distance, or $\mathcal{L}_2$ distance, between two vectors $x, y \in \mathbb{R}^d$, is computed by the number of distinct tokens at the corresponding position, which is formally formulated as:\vspace{-1em}
    $$dist_{Euclidean}(x, y) = \sum_{i=1}^{d}(x_i-y_i)^2.$$
    \label{definition: l2 distance}
\end{definition}



\subsection{LSH and Product Quantization}

\htitle{Locality-Sensitive Hashing.} Locality-Sensitive Hashing (LSH) is a prestigious vector index, which is widely used approximate nearest neighbor search. The core property of LSH is that closer points have larger probability of collision than those distant points. Formally, the property is defined as following:

\begin{definition}
    \cite{DBLP:conf/icde/TianZZ22} Given a distance $r\geq0$ and an approximation ratio $c > 1$, a LSH function family $\mathcal{H} = \{h: \mathbb{R}^d \rightarrow \mathbb{R}\}$ is considered $(r, cr, p_1, p_2)$-sensitive if $\forall o_1, o_2 \in \mathcal{D}$:
    \begin{enumerate}
        \item if $||o_1, o_2|| \leq r$, then $Pr[h(o_1) = h(o_2)] \geq p_1$,
        \item if $||o_1, o_2|| > c\cdot r$, then $Pr[h(o_1) = h(o_2)] \leq p_2$.
    \end{enumerate}
\end{definition}

Different distance functions, such as $\mathcal{L}_p$, Jaccard distance, uses different LSH functions to represent the distance proximity relations. In this work, we focus on the Euclidean space. A typical LSH family for Euclidean space in E$2$LSH is defined as follows:
$$h_{a,b}(o) = \lfloor\frac{\vec{a} \cdot \vec{o} + b}{W}\rfloor\label{definition: lsh},$$ where $\vec{o}$ is the vector representation of a data point in $\mathcal{D}$, $\vec{a} \in \mathbb{R}^d$ is a $d$-dimensional vector whose entries are sampled independently from $2$-stable distribution (standard normal distribution), $W$ is a predefined integer, and $b \in \mathbb{R}$ is uniformly sampled from $[0, W]$.

\htitle{Basic LSH Indexing.} In this part, we will go through the basic index construction process of locality sensitive hashing in approximate nearest neighbor search. Firstly, we have a family of locality sensitive hashing functions $\mathcal{H} = \{h_1, h_2, ..., h_N\}$, then $k$ numbers of functions will be sampled from $\mathcal{H}$, which forms a composite hash functions $g$ and the hash value of vector $v$ is computed as $g(v) = (h_1(v), h_2(v), ..., h_k(v))$. And this process constructs a single hash table, where vectors are assigned a unique hash code, based on distance proximity. Typically, in context of ANNS in Euclidean space, a LSH index contains $L$ hash tables, where each uses $K$ hash functions, and form a $(K-L)$ LSH scheme.

After understanding the basic idea of how locality sensitive hashing index works, now let's define two extended concepts that will be essential for understanding the core of our framework.

\begin{definition}[\textbf{Central Bucket $\mathcal{B}_{central}$}]
    In a hash table, given a family of LSH functions: $\{h_1, h_2, ..., h_N\}$, a central bucket $\mathcal{B}_{central}$ of a data point $x \in \mathbb{R}^d$ is defined as the hash bucket that the query vector $x\in \mathbb{R}^d$ directly hashed to. The hash code of $\mathcal{B}_{central}$ is: $$\mathcal{B}_{central} = (h_1(x), h_2(x), .., h_N(x)).$$
\end{definition}

\begin{definition}[\textbf{Hamming Distances}]
    The hamming distance, between two vectors $x, y \in \mathbb{R}^d$, is computed by the number of distinct tokens at the corresponding position, which is formally formulated as:\vspace{-1em}
    $$dist_{hamming}(x, y) = \sum_{i=1}^{d}(x[i] == y[i])$$
\end{definition}

\begin{definition}[\textbf{$k$-Step Neighbor of $\mathcal{B}_{central}$}]
    Given a central bucket $\mathcal{B}_{central}$, the $k$-step neighbor of the central bucket is a set of hash buckets, denoted as $\mathcal{N}_k = \{\mathcal{N}_{k1}, \mathcal{N}_{k2}, ..., \mathcal{N}_{ki}\}$, where the hash code of each $\mathcal{N}_{ki}$ is a hash bucket satisfying $$dist_{hamming}(\mathcal{B}_{central}, \mathcal{N}_{ki}) == k$$
\end{definition}

\htitle{Product Quantization.} Product quantization\cite{DBLP:journals/pami/JegouDS11} is a popular index used in approximate nearest neighbor search in high dimensional space, whose core idea is to compress vectors of full length into compacted form to make it memory-friendly. Now, we briefly demonstrate the process of product quantization and explain the part that will be utilized in our work. Technically speaking, a vector $x \in \mathbb{R}^d$ will be firstly split into $M$ subvectors $x = (x_1, x_2, ..., x_M)$, where each of these subvectors is of dimension $\frac{d}{M}$ and $M$ divides $d$ should be satisfied; then all vectors in the dataset will be divided in $M$ subspace, where each subspace consists of all subvectors of the corresponding index. For each of the $M$ subspace, PQ conducts clustering algorithm, like KMeans, to form $K$ numbers of centroids of the subspace, where each (sub)vector is assigned with the closest centroid. After deriving the centroids and point assignment of each subspace, each vector can be represented as codebook vector of length $M$, like $x = (3, 2, ..., 8)$, where each of the $M$ value represent the identification of the closest centroid of that subspace. Conversely, vectors can also be represented as the concatenation of the series of centroids in each of the $M$ subspaces. 

In PQ, as each vector is compressed and represented by a codebook, the distance computation between different vectors can be more efficient and facilitated. Overall, there are two schemes of distance computation in product quantization, referred to as symmetric distance computation (SDC) and asymmetric distance computation (ADC). Specifically, SDC will compute the distance between two vectors based on their quantized vector, which is mathematically represented as:\vspace{-1em}
$$\hat{d}(x, y) = d(q(x), q(y)) = \sqrt{\sum_{j}{d(q_j(x), q_j(y))^2}},$$
whose advantage is that, in each subspace, the distance between any two codebook vectors can be precomputed, tackling the challenge that the queries are unknown. Despite the convenience and efficiency in computation, $SDC$ suffers from a low accuracy in distance approximation, and it can rarely be used in tasks requiring high precision. On contrast, ADC improves the shortcoming of $SDC$ by not quantizing the query vector, which means that the query vector remains with full precision. Mathematically, $ADC$ is computed as: $$\hat{d}(x, y) = d(x, q(y)) = \sqrt{\sum_{j}{d(x_j, q_j(y))^2}}.$$
The tradeoff is that we will not be able to pre-compute a universal distance table for all queries, as queries are unknown, which slightly damages the efficiency. However, we will still be able to compute the distance between each subvectors of the query and codebook vectors of each subspace, so that the further distance computations can still refer to the distance table. Considering the precision requirement in cardinality estimation, as well as approximate nearest neighbor search, $ADC$ will be utilized in our work and will be further discussed in the optimization section.

\section{Existing Solutions}
\label{sec: related work}
The cardinality estimation problem is a classic problem in database for decades, where numerous solutions have been proposed in relational database\cite{DBLP:journals/pvldb/HanWWZYTZCQPQZL21, DBLP:journals/pvldb/YangKLLDCS20, DBLP:journals/pvldb/YangLKWDCAHKS19, DBLP:conf/sigmod/DeshpandeGR01, DBLP:journals/vldb/GunopulosKTD05, DBLP:conf/sigmod/GetoorTK01, DBLP:journals/pvldb/ShetiyaTK020, DBLP:journals/pvldb/KieferHBM17}. In vector database, the cardinality estimation problem is to estimate the number of qualified points in a similarity search. We will review several major works that are closely related to this research.

\htitle{Sampling Method.} The most naive and trivial approach for solving a cardinality problem for similarity search is uniform sampling, whose convention is to uniformly sample $1\%$ or $10\%$ of the dataset to derive the estimation for the entire dataset. And the method is commonly used as the competitor in most of the recent works on the problem\cite{DBLP:conf/sigmod/Sun0021, DBLP:conf/icml/WuCN18, 10.14778/3712221.3712224}. While sampling-based approaches offer a straightforward means to estimate the cardinality, they suffer from two major limitations. First, the resulting estimates are often coarse and lack accuracy. Second, their efficiency is poor: to achieve accuracy comparable to other methods, sampling based techniques require the evaluation over a substantially larger amount of points, leading to significant efficiency degradation.

\htitle{Hyperplane LSH Approach\cite{DBLP:conf/icml/WuCN18}.} It proposed a density estimator using the locality sensitive hashing and importance sampling, which \textit{emphasized on the estimation in angular space}. It firstly partitions the dataset with hyperplane LSH, then, it uniformly samples from promising buckets, according to the hamming distance of different buckets. Formally, the aforementioned process is formalized with a random variable: $Z = \frac{\sum_{k=1}^K C_{q}^{k}(I)}{K \cdot p(x)}$, where $C_{q}^{k}(I)$ is the number of points in buckets at distance $I$ and $K$ represents the number of hash tables. Most importantly, the normalization factor $p(x)$ represents the collision probability that some point $x$ lands in a bucket at hamming distance $I$ from $q$ with randomly chosen hash functions. Lastly, the random variable will be sampled for $S$ times $Z_1, Z_2, ..., Z_S$ and estimation is taken as the average.


\htitle{SimCard\cite{DBLP:conf/sigmod/Sun0021}}.
SimCard is a deep neural network based estimator for high-dimensional spaces that adopts a global–local structure for cardinality estimation. Specifically, it applies K-Means to partition the dataset and employs PCA for dimensionality reduction. For each cluster, a local estimation model is trained to predict the query cardinality within that partition. However, evaluating a query across a large number of local models (e.g., $100$) can be computationally expensive. To mitigate this, a global model is introduced to identify the most promising local regions for cardinality estimation. The models take as input a set of features, including the query vector, the distance threshold, and some properties of the dataset.

The framework is novel in its global–local architecture; however, it also inherits several common drawbacks of learning-based approaches. In particular, the performance of the estimator is highly dependent on the quantity and quality of training data, and insufficient or low-quality data can significantly degrade accuracy. Moreover, in practice, the framework comprises hundreds of neural networks serving as local estimators, which makes it impractical and inefficient to support large-scale dataset updates.


\htitle{SRCE / MRCE\cite{10.14778/3712221.3712224}}. 
SRCE and MRCE represent recent advances in cardinality estimation for high-dimensional spaces. SRCE is a simple estimator that exploits fundamental properties of similarity queries, whereas MRCE incorporates machine learning techniques to enhance prediction accuracy. Both methods leverage knowledge of the existing dataset, referring to as \textit{reference objects}, operating under the assumption that distance function of a given query point may resemble that of certain points already present in the dataset.


SRCE estimates cardinality using a single reference object, generating a pool of reference objects while balancing pool size and diversity. During online estimation, SRCE identifies the reference point whose distance function most closely matches that of the query. Computing the distance function online is time-consuming; to address this, SRCE employs a vector index to pre-compute distance functions during the offline phase. Despite this efficiency improvement, the need for pre-computation makes SRCE inherently query-dependent, which is a notable limitation of the algorithm.


MRCE employs multiple reference objects to estimate cardinality, aiming to reduce SRCE’s reliance on a vector index for pre-computing distance functions. With multiple references, MRCE trains a deep neural network to evaluate the contribution of each reference, and the weighted sum of these contributions produces the final cardinality estimate. In terms of network architecture, the model takes as input a data point $p_i$, a reference object $r_i$, the distance between $p_i$ and $r_i$, and the distance threshold. During the first training phase, an encoder–decoder model is trained to featurize $p_i$ and $r_i$, and the resulting embeddings are concatenated with the remaining distance features to compute the final weights for each reference object, which are then combined to produce the cardinality estimate. MRCE reduces its reliance on the vector index; however, obtaining a sufficient amount of diverse training data to achieve satisfactory estimation performance is time-consuming. Moreover, while SRCE and MRCE are novel in leveraging existing dataset knowledge, in practice, dozens of values must be pre-computed to ensure both efficiency and accuracy.

\section{Our Framework}
\label{sec: framework}
\subsection{Overview}
The core idea is to partition the dataset using locality-sensitive hashing (LSH), which enables efficient pruning of unpromising points while quickly identifying candidates in close proximity to a given query (Section \ref{subsec: Dataset Partitioning}). Beyond the LSH index, we introduce a neighboring-based probing strategy (Section \ref{subsec: Neighboring-based Probing Strategy.}) to adaptively adjust the number of points explored depending on the magnitude of distance threshold (Section \ref{Adaptive Bucket Probing: An Overview}), thereby enhancing the efficiency of estimation. Furthermore, we introduce two extra optimizations. First, we use the progressive sampling to reduce the number of required distance computations while providing strong probabilistic guarantees on accuracy (Section \ref{subsec: progress sampling}). Second, we leverage asymmetric distance in product quantization\cite{DBLP:journals/pami/JegouDS11} to further accelerate distance computation in high dimensional space (Section \ref{subsec: product quantization}).

The pseudocode of the neighboring probing strategy with progressive sampling is demonstrated in \textbf{Algorithm \ref{algorithm: neighboring based probing}}, where $f\_neighbor$ (Line \ref{algoLine: f_neighbor}): \textbf{Algorithm \ref{algorithm: f_neighbor with progressive sampling}} and $f\_central$ (Line \ref{algoLine: f_central}): \textbf{Algorithm \ref{algorithm: central bucket}}.

In addition to static dataset, our framework also supports data updates, where each component of our framework can be updated easily, even with large scale of new data. And we provide the algorithm for updating the framework in Section \ref{sec: dynamic data}.

\subsection{Dataset Partitioning}
\label{subsec: Dataset Partitioning}

Dataset partitioning, or segmentation, is a common preprocessing strategy in cardinality estimation for high-dimensional vectors. Its necessity can be summarized as follows:
\begin{enumerate}[left =0pt]
    \item Modern vector datasets often contain millions of points, making direct cardinality estimation over the entire space both challenging and prone to significant errors \cite{DBLP:conf/sigmod/Sun0021}. Consequently, partitioning the dataset into smaller subsets is beneficial for improving estimation accuracy and efficiency.
    \item In cardinality estimation, points that are sufficiently distant from the query cannot serve as viable candidates, making their exploration unnecessary. So, partitioning the dataset is advantageous for focusing on the most promising candidates, thereby improving efficiency by pruning unpromising points.
\end{enumerate}

Our work utilizes the locality-sensitive hashing (LSH) to partition the dataset. As the work focuses on the Euclidean space, the LSH function is naturally selected as the E$2$LSH\cite{e2lsh}. In addition, in order to support the angular space, the framework can be easily adopted with hyperplane LSH, which is also the choice of a similar work under angular space\cite{DBLP:conf/icml/WuCN18}.

\subsection{Neighboring-based Probing Strategy.}
\label{subsec: Neighboring-based Probing Strategy.}

\htitle{Design Objective.} To tackle the cardinality estimation of similarity queries in high dimensional space, we aim to develop a \textit{distance threshold aware} bucket probing strategy, which takes advantage of the information (like selectivity) of those explored hash buckets, to adaptively adjust the number of buckets to be explored. In terms of efficiency, the goal is that our probing strategy will be able to give an accurate estimation while probing a tiny proportion of the dataset to minimize the number of distance computation required, which is considered as an expensive part during the online estimation.

To achieve such objective, we are motivated by the multi-probe LSH\cite{DBLP:conf/vldb/LvJWCL07} in approximate nearest neighbor search (ANNS), which claims that, given a central bucket $\mathcal{B}_{central}$ in a hash table, its neighboring buckets that are at few distance away (measured by the distance between hash codes in hamming space), are also probable to contain the neighbors to the query. In this work, we further explore and utilize this property to tackle the estimation problem, and we will provide detailed reasoning and justification for the design of each component of our framework. 

\htitle{Challenges: Numerous Hash Buckets.}
In locality sensitive hashing (LSH), in order to achieve a decent performance in space partitioning while capturing distance proximity among millions of points, it is a common practice to apply multiple LSH functions to better capture the distance distribution\cite{DBLP:conf/icde/TianZZ22, 10184454}. However, the usage of multiple LSH functions give rise to a magnificent amount of hash buckets in the hash table.
\begin{example}
Given one hash table, we use $10$ hash functions, where each of the function produces around $4$ different values, and theoretically, there are around $4^{10} = 1,048,576$ distinct hash buckets being created in the hash table. Furthermore, if $14$ hash functions are used, there will be around $4^{14} = 268,435,456$ buckets. 
\end{example}
Therefore, it will be extremely inefficient to probe the neighbors at bucket level for the following reasons: firstly, it is inefficient to probe with single buckets, where the neighbor look-up / indexing time will incur large latency during online estimation and it will be infeasible of estimating a meaningful number of buckets to be explored; secondly, according to our preliminary study, the number of points at each bucket can be highly imbalanced, which means numerous buckets might contain very few points (like $3$ or $10$) and thus degrades the locality information of the dataset.

\htitle{Our Approach.}
Considering the disadvantages of directly probing across numerous hash buckets, we propose a \textit{$k$-Step Neighboring-based Probing strategy}, whose fundamental notion is from the multi-probe locality sensitive hashing. Firstly, let's re-clarify the notion of a $k$-step neighbor $\mathcal{N}_k$ of a hash bucket $\mathcal{B}_{central}$, which is a set of hash buckets, whose hash codes are at $k$ distance away from the $\mathcal{B}_{central}$ in the hamming space. Now, our general probing framework is that, given a query vector $q$, we firstly estimate its central hash bucket $\mathcal{B}_{central}$ (Line \ref{algoLine: f_central}), then we explore those neighboring buckets $\mathcal{N}_1$, $\mathcal{N}_2$, .., $\mathcal{N}_{K'}$ (Lines \ref{algoLine: neighbor_start}--\ref{algoLine: neighbor_end}) and terminates if either met:
\begin{itemize}[left=0pt]
    \item number of points explored meets maximum value (Lines \ref{algoLine: max point reach start}--\ref{algoLine: max point reach end})
    \item global probing termination flag is triggered (Lines \ref{algoLine: PTF start}--\ref{algoLine: PTF end}),
\end{itemize}
where $K' \leq K$ and the maximum of $K$ is the number of hash function (\textbf{Algorithm \ref{algorithm: neighboring based probing}}).


\begin{algorithm}[t]
    \caption{Neighboring Based Probing}
    \label{algorithm: neighboring based probing}
    
    \textbf{INPUT:} Query: $(x \in \mathcal{R}^d ,\tau \in \mathcal{R})$, LSH Index: $\mathcal{I}$; Fast Neighbor Lookup Table $\mathcal{P}$, estimation function: $f$; \\
    
    \textbf{OUTPUT:} Estimated Cardinality $|\mathcal{A}|$ \\
    
    $|\mathcal{A}| \gets 0$; \\
    $nVisted \gets 0$; (diff from actual computation) \\

    PTF $\gets none$ (global probe terminate flag);\\
    
    $hashcode \gets \mathcal{I}.computeHashCode(x)$; \\

    $|\mathcal{A}| += f_{central}(x, \tau, \mathcal{I}, hashcode)$;\label{algoLine: f_central} \\

    $nHashFuncs \gets \mathcal{I}.n\_lsh\_funcs$; \\

    \For{$nDegree \in range(1, nHashFuncs)$\label{algoLine: neighbor_start}}{
        
        \If{$nVisted \geq maxVist$\label{algoLine: max point reach start}}{
            break;\\
        }\label{algoLine: max point reach end}

        nDeg\_card, PTF $= f_{neighbor}(nDegree, hashcode, x, \tau, \mathcal{P})$;\label{algoLine: f_neighbor} \\

        $|\mathcal{A}| += \text{nDeg\_card}$; \\

        \If {PTF\label{algoLine: PTF start}}{
            break; \\
        } \label{algoLine: PTF end}
        update $nVisted$;\\
    }\label{algoLine: neighbor_end}
    
    \Return $|\mathcal{A}|$;\\
\end{algorithm}

\begin{example}
    Given a query vector $q \in \mathbb{R}^d$, it is easy to compute the hash code of the central bucket $\mathcal{B}_c = [0, 2, 1, 3]$, where we use one hash table with four hash functions. In our hash table, there are a set of distinct hash buckets:

    \begin{itemize}
        \item $\mathcal{N}_1$: $[1, 2, 1, 3], [0, 2, 1, 4]$
        \item $\mathcal{N}_2$: $[2, 3, 1, 3], [0, 1, 2, 3], [1, 2, 1, 4]$
        \item $\mathcal{N}_3$: $[0, 3, 2, 4], [1, 2, 2, 2], [1, 1, 0, 3]$
        \item $\mathcal{N}_4$: $[1, 1, 2, 2],$
    \end{itemize}
    where each hash bucket contains points sharing the hash code, and we sort these hash codes by its distance to the hash code of central bucket. In our probing strategy, points in the same $\mathcal{N}_k$ are grouped together as a whole. Then, our strategy is to derive the estimation with $\mathcal{N}_k$ by increasing $k$ until certain stopping conditions are met. 
\end{example}

\subsection{Adaptive Bucket Probing}
\label{Adaptive Bucket Probing: An Overview}

\htitle{Challenges: Adaptiveness of Threshold.} 
Designing an efficient and accurate probing strategy is inherently challenging. In the context of cardinality estimation, exploring too many or too few points can result in poor efficiency or inadequate accuracy, respectively, making it crucial to balance this trade-off. As previously discussed, a similarity search query includes not only a vector point but also a distance threshold. The threshold can be large enough to encompass over $10\%$ of the dataset, necessitating the exploration of many neighbors or buckets, or it can be very small, covering only a few points—e.g., $2$ or $5$—where examining a single bucket suffices. Furthermore, for datasets with complex distance distributions, exploring too few buckets can cause a significant drop in accuracy, as the cardinality may increase dramatically with even a slight increase in the distance threshold. Therefore, determining the appropriate number of points to explore is of critical importance.

Given this challenges, our probing algorithm is expected to be \textit{adaptive} to distance thresholds at different magnitude, which means that our algorithm is expected to automatically adjust the number of points to be explored based on the magnitude of distance threshold, where more points will be explored for larger thresholds, while less points for smaller thresholds. What's more, as previously discussed, threshold is known only at the online estimation phase and there is no such a golden rule to determine if the threshold is large or small, which even differs between different points in the dataset, as well as datasets with various distributions. In short, our algorithm is expected to efficiently determine the number of points to explore during the online estimation. Now, let's discuss our general design of adaptive prober in detail.

\htitle{A Naive Approach.} As aforementioned, one of the objective is to achieve accurate estimation while exploring a tiny proportion (like $1\%$) of dataset. A straightforward approach is to explore from the central bucket $\mathcal{B}_{central}$, and then consecutively proceed to $k$-th neighbor $\mathcal{N}_k$ at farther distances, until $x\% \cdot |D|$ numbers of points are explored. Considering points in the same hash buckets have similar proximity distribution, instead of explicitly computing distances of all points from the query, we uniformly sample $\frac{1}{x}\% \cdot |\mathcal{N}_k|$ points to compute the distance. In order to enable the prober aware of the magnitude of distance threshold, we determine if further neighborhood $\mathcal{N}_{k+1}$ should be explored based on the selectivity of the current neighborhood, where the termination condition is triggered if selectivity is under some threshold.

\textit{Comment:} The uniform sampling and selectivity-based termination strategy partially tackles the challenges in efficiency and being adaptive to deal with thresholds at various magnitudes. However, the sampling strategy and termination condition is highly heuristic, which cannot provide any rigorous theoretical guarantee regarding the confidence of error bound in estimation; and there is still noticeable error in estimation.

To reduce the estimation error and reduce the number of point computations, we propose our first optimization that utilizes progressive sampling, with theoretical guarantee, to adaptively adjust the number of points for exploration within a neighbor $\mathcal{N}_k$, which is discussed in Section \ref{subsec: progress sampling}. As dimensionality increases, the complexity of distance computation increases. In order to mitigate the impact from the dimensionality of vectors, the second optimization is to apply asymmetric distance computation (ADC) to estimate the real distance between two points, which is discussed in Section \ref{subsec: product quantization}.

\subsection{Adaptive Progressive Sampling with Guarantee}
\label{subsec: progress sampling}

Progressive sampling  is an adaptive sampling method, comparing with uniform sampling. The motivation is that some neighborhoods $\mathcal{N}_k$ might contain a large amount of points, where a small proportion of them is representative enough to reflect its distribution and sampling more points is unnecessary. What's more, progressive sampling allows sampling for multiple times, which better avoids outlier cases and further improves accuracy.

Besides the progressive sampling, we design our algorithm to be adaptive to thresholds of different magnitudes, where we design two termination conditions, the first of which terminates sampling with greater sample size of the current $\mathcal{N}_k$ and the second of which terminates the entire neighbor probing process. Both termination conditions can be formulated as a function of the sampling parameters and selectivity in neighbor $\mathcal{N}_k$, and we will explain these concepts in detail with following section and \textbf{Algorithm \ref{algorithm: f_neighbor with progressive sampling}}.

Given a $k$-th step neighbor containing $N$ points, we define a sampling schedule as $s_1, s_2, s_3, .., s_T$, where $s_{i+1} = 2 \cdot s_i$ (Line \ref{algoLine: double sampling rate}), and we denote the number of sampled points to be $w = s_i \cdot N$. In practice, we also set an upper bound for the maximum sampling rate as $s_{max}$, where $s_T \leq s_{max}$, to avoid exploring infinitely (Line \ref{algoLine: max sampling rate control}). To achieve a theoretical bound for the confidence of estimation, we derive the upper bound of error as (Line \ref{algoLine: upper bound}):
$$\mu_{upper} = (\sqrt{\hat{p} + \frac{a}{2w}}+ \sqrt{\frac{a}{2w}})^2,$$
and derive the lower bound as follows (Line \ref{algoLine: lower bound}):
$$\mu_{lower} = max\{0, \left(\sqrt{\hat{p} + \frac{2a}{9w}} - \sqrt{\frac{a}{2w}}\right)^2 - \frac{a}{18w}\}.$$
where $w$ denotes the number of sampled points. And $\hat{p}$ (Line \ref{algoLine: p_hat}) represent the selectivity of the current round of sampling, formally, it is defined as: 
$$\hat{p} = \frac{w'}{w},$$
where $w'$ represent the number of points that are qualified within distance threshold, which is evaluated iteratively (Lines \ref{algoLine: iter point start}--\ref{algoLine: iter point end}), and $w = s_i \cdot N$ refers to sample size (Line \ref{algoLine: sample size assign}).

Additionally, in Lines \ref{algoLine: iter point start}--\ref{algoLine: iter point end}, $dist_{\mathcal{L}_2}$ computes the distance between two points in $\mathcal{L}_2$ space. In our framework, $dist_{\mathcal{L}_2}$ can be calculated as Definition \ref{definition: l2 distance} or approximate it with asymmetric distance in PQ\cite{DBLP:journals/pami/JegouDS11}, which will be discussed in Section \ref{subsec: product quantization}.

Now, let $\epsilon$ denote the error bound parameter $\epsilon$. For example, $\epsilon=0.0001$. In our probing scheme, we terminate from sampling more points at the current neighborhood if the first condition $(1)$ met, which means we are already confident regarding the error bound of our estimation (Lines \ref{algoLine: terminate following sampling start}--\ref{algoLine: terminate following sampling end}) and there is no need to increase the sample size; and the second condition $(2)$ is even stronger, which indicates that there is no need to explore neighbors at further (hamming) distances (Lines \ref{algoLine: terminate all probing start}--\ref{algoLine: terminate all probing end}): 
\begin{align}
\mu_{\text{upper}} - \hat{p} &\leq \epsilon \wedge \hat{p} - \mu_{\text{lower}} \leq \epsilon \tag{1} \\
\mu_{\text{upper}} &< \epsilon \tag{2}
\end{align}

Lastly, we provide reasoning of the termination conditions. Firstly, we assume that, the probability of failure (larger than the error bound) for our estimation to be $PR_{fail}$ (say $0.1\%$), which means that the probability of success is $PR_{success}= 1 - PR_{fail}$ and we denote the constant $a = \ln(\frac{1}{PR_{fail}})$. And we may interpret two stopping conditions $(1)(2)$ as the upper and lower bound of the error for our estimation, which provides a guarantee with confidence $PR_{success}$. The formula is deduced with \textit{Chernoff bound}.


\begin{algorithm}[t]
    \caption{$f_{neighbor}$: Estimate the cardinality of $\mathcal{N}_k$ with Progressive Sampling}
    \label{algorithm: f_neighbor with progressive sampling}
    
    \textbf{INPUT:} nDegree (distance w. $\mathcal{B}_{central}$), Query: $(x \in \mathcal{R}^d ,\tau \in \mathcal{R})$, LSH Index: $\mathcal{I}$; Neighbor Lookup Table $\mathcal{P}$; \\

    \textbf{PARAMS:} Initial Sampling Rate: $s_1$, Max Sampling Rate: $s_{max}$, Error bound: $\epsilon$; Confidence Param: $a=ln(1000)$\\
    
    \textbf{OUTPUT:} Estimated Cardinality $|\mathcal{A}|$ \\

    $|\mathcal{A}| \gets 0$; \\

    $hashcode \gets \mathcal{I}.computeHashCode(x)$; \\

    $\mathcal{C} \gets \mathcal{P}.find(nDegree, hashcode)$; (Hashcode of Neighbors); \\
    
    $\mathcal{O} \gets \mathcal{I}.retrieve(C)$; (nDegree Neighbor IDs); \\

    $CSR = s_1$; (current sampling rate); \\

    $PTF \gets false$; (global probing termination flag); \\

    $Q_{all}, Q_{qualified} \gets 0, 0$; \\

    \While{$CSR \leq s_{max}$\label{algoLine: max sampling rate control}}{

        $\mathcal{O}' \gets sampler(\mathcal{O}, CSR)$; (sampled points) \\

        $w, w' \gets |\mathcal{O}'|, 0$; (current sample size, sample qualified)\label{algoLine: sample size assign}\\

        \For{$p \in \mathcal{O}'$\label{algoLine: iter point start}}{
            $distance_{curr} \gets dist_{\mathcal{L}_2}(x, p)$; \\
            \If{$distance_{curr} \leq \tau$}{
                $w' ++$; \\
            }
        }\label{algoLine: iter point end}

        $\hat{p} \gets \frac{w'}{w}$\label{algoLine: p_hat}; \\

        $\mu_{upper} = (\sqrt{\hat{p} + \frac{a}{2w}}+ \sqrt{\frac{a}{2w}})^2$; \label{algoLine: upper bound}\\

        $\mu_{lower} = max\{0, \left(\sqrt{\hat{p} + \frac{2a}{9w}} - \sqrt{\frac{a}{2w}}\right)^2 - \frac{a}{18w}\}$; \label{algoLine: lower bound}\\

        $Q_{all} += w$;\\
        
        $Q_{qualified} += w'$; \\

        \If{$\mu_{upper} < \epsilon$\label{algoLine: terminate all probing start}}{
            $PTF \gets true$; \\
            break; \\
        }\label{algoLine: terminate all probing end}

        \If{$\mu_{upper} - \hat{p} \leq \epsilon \wedge \mu_{lower} - \hat{p} \leq \epsilon$\label{algoLine: terminate following sampling start}}{
            break; \\
        }\label{algoLine: terminate following sampling end}

        $CSR \gets 2 \cdot CSR$;\label{algoLine: double sampling rate} \\
    }

    $|\mathcal{A}| = |\mathcal{O}| \cdot {\frac{Q_{qualified}}{Q_{all}}}$; \\

    \Return $|\mathcal{A}|$, $PTF$; \\
    
\end{algorithm}

\begin{algorithm}[!t]
    \caption{$f_{central}$: Estimate in Central Bucket}
    \label{algorithm: central bucket}
    The algorithm describes how the estimation is executed in the central bucket: $\mathcal{B}_{central}$.
    
    The algorithm is in naive brute force manner, where it iteratively computes the distance between query and point, and the point is counted if the distance is no greater than the query threshold.
\end{algorithm}

\subsection{PQ-based Distance Estimation}
\label{subsec: product quantization}

As previously introduced, product quantization\cite{DBLP:journals/pami/JegouDS11} is an efficient and memory-friendly vector index that is popular in approximate nearest neighbor search. In our work, we leverage the asymmetric distance computation (ADC) in PQ to further accelerate the distance computation, which is a costly part in the online estimation process. 

\begin{algorithm}[t]
    \caption{PQ-ADC: Construct Fast Lookup Table}
    \label{algorithm: pq adc table construct}
    \textbf{INPUT:} PQ Index: $\mathcal{Q}$, Query Vector $x \in \mathbb{R}^d$;\\
    \textbf{PARAMS:} $M$: number of subspaces, $K$: number of clusters;\\
    \textbf{OUTPUT:} Query Specific Fast Lookup Table $\mathcal{T}$;\\

    Initialize fast lookup table $\mathcal{T}$; \\

    $x_1, x_2, ..., x_M = divide\_subspace(x)$; \\

    \For{$spID \in [1, 2, ..., M]$}{
        \For{$cID \in [1, 2, ..., K]$}{
            Retrieve the centroid vector: $vec^{spID}_{cID}$; \\
            Compute the distance: $dist = l2\_dist(x_{spID}, vec^{spID}_{cID})$;\\
            $\mathcal{T}[spID][cID] = dist$; \\
        }
    }
    \Return {$\mathcal{T}$}; \\
\end{algorithm}

When query arrives, we firstly compute the distance between each (sub)vector of the query with the centroid of codebook vectors, where we store all these values in a fast lookup table $\mathcal{T}$ (\textbf{Algorithm: \ref{algorithm: pq adc table construct}}); then, for all future distance computation, we directly refer to the table for better efficiency (\textbf{Algorithm \ref{algorithm: pq adc compute}}). It is expected that ADC accelerates distance computation, especially for high dimensional vectors, while introducing little estimation errors.

\begin{algorithm}[t]
    \caption{PQ-ADC: Distance Estimation}
    \label{algorithm: pq adc compute}
    \textbf{INPUT:} PQ Index: $\mathcal{Q}$, Fast Lookup Table: $\mathcal{T}$, Point ID: $pid$ (data points whose distance to be computed with query); \\
    \textbf{PARAMS:} $M$: number of subspaces, $K$: number of clusters;\\
    \textbf{COMMENT:} Refer to our GitHub repository for a more compiler friendly and efficient implementation.; \\

    $dist_{adc} \gets 0$; \\

    \For{$spID \in [1, 2,..., M]$}{
        $dist_{adc} += \mathcal{T}[spID][\mathcal{Q}.codebook.pid]$; \\
    }

    \Return {$dist_{adc}$}; \\
\end{algorithm}

\subsection{Efficient Bucket Neighbor Look-Up}
\label{subsec: Bucket Neighbor Look-Up}

The estimation is expected to be efficient. However, in our neighboring based probing paradigm, one of the efficiency bottleneck is to efficiently retrieve all the hash buckets / points, whose hash code is at certain distance from $\mathcal{B}_{central}$. The challenge is that a hash table usually contains magnificent amount of buckets, like over $100,000$, and it is costly to iteratively retrieve them during the online estimation. A natural solution is to pre-compute it offline. Specifically, we construct a neighbor lookup table $\mathcal{P}$, where each index (position) of $\mathcal{P}$ stores the neighboring information of corresponding hash code in $\mathcal{C}$. At each index $i \in [0, \mathcal{C}.size)$ of $\mathcal{P}$, a dictionary structure is used, where we use the distances between hash codes as the key and the hash code id will be stored as the value. For example, $\mathcal{P}[i][k]$ returns the index of all hash codes, which are $k$ distance away from the hash code $\mathcal{C}[i]$. The process is described in \textbf{Algorithm \ref{algorithm: fast neighbor lookup}}.

Additionally, according to our preliminary study, we observe that it is unnecessary to store the neighbor information with large distance, which might not be accessed in our probing scheme. To reduce the cost of storage, we store neighbor with distance no greater than $M$ at each index.

\begin{algorithm}[t]
    \caption{Construct Neighbor Lookup Table}
    \label{algorithm: fast neighbor lookup}
    \textbf{INPUT:} $\mathcal{C}$: Array of unique hash codes;\\
    \textbf{PARAMS:} $M$: distance greater than it will not be stored; \\
    \textbf{OUTPUT:} $\mathcal{T}$: Efficient Neighbor Lookup Table;\\

    \For{$i \in [0, \mathcal{C}.size)$ }{
        \For{$j \in [0, \mathcal{C}.size)$}{
        $d = dist_{hamming}(\mathcal{C}[i], \mathcal{C}[j])$; \\
        \If{$0 < d \leq M$}{
            $\mathcal{T}[i].\text{update}\{(d, j)\}$; \\
        }
    }
    \Return{$\mathcal{T}$}; \\
    }
\end{algorithm}


\section{Dynamic Data Updates}
\label{sec: dynamic data}
Supporting data updates is practical but challenging in modern vector database, where it requires dynamically updating the constructed index as rebuilding from scratch is inefficient. Technically, the problem will become more challenging for large scale data updates, where the data distribution will change significantly. 

Existing frameworks are primarily learned frameworks\cite{DBLP:conf/sigmod/Sun0021, 10.14778/3712221.3712224}, where they train DNNs with various types of labeled data, and the model will be used for the cardinality estimation. In terms of data update, current frameworks often applies the following strategies:
\begin{enumerate}[left=0pt]
    \item Directly use the original trained model for estimation.
    \item Re-compute labels of training data, and re-train the model.
\end{enumerate}

However, both data update scheme suffers from some disadvantages. For the first approach, the model performance degrades little when few new data is added, but the performance degrades significantly as the scale updates increase, because the training labels are no longer valid under the updated dataset. The second approach is rigorous, however, it is time-consuming to re-compute the training labels and re-train the deep neural networks for each update. For example, in SimCard \cite{DBLP:conf/sigmod/Sun0021}, we need to firstly re-compute the cluster-wise training labels, and then we need to re-train over $100$ deep neural networks (each data partition has a local DNN for estimation) to achieve satisfying results; in \cite{10.14778/3712221.3712224}, it re-generates all or partial of training labels, which contain magnificent amount of intermediate data, to retrain the model to support updates. 

Fortunately, leveraging the advantage of locality sensitive hashing, we are able to efficiently support large scale of data updates without significant accuracy degradation with the updated framework. Overall, there are three major components of the framework to be updated: 1. locality sensitive hashing index $\mathcal{I}$; 2. product quantization index $\mathcal{Q}$ (optional); 3. neighboring bucket lookup table $\mathcal{P}$. Now, I will provide algorithm for updating each of component, and performance of updated framework will be reported in Section: \ref{sec: dynamic data}.

\htitle{Update LSH Index.} \textbf{(Algorithm \ref{algorithm: dynamic lsh update})} We firstly iteratively compute the hash code of newly added points with original hash functions. However, a negligible but beneficial step is to re-calculate the value of $W$ (refer to Definition: \ref{definition: lsh}) of the LSH function, which is determined by the minimum and the maximum value of all hash codes. If $W$ is not properly updated, quality of updated hash table degrades.

\begin{algorithm}[t]
    \caption{Dynamic Updates: LSH Index}
    \label{algorithm: dynamic lsh update}
    \textbf{INPUT:} Existing LSH Index: $\mathcal{I}$, New Points: $\mathcal{V} = \{v_i\}$;\\
    \textbf{OUTPUT:} Updated LSH Index: $\mathcal{I}$; \\

    $\mathcal{H} \gets \mathcal{I}.lsh\_functions$; (division excluded) \\

    $HashCodes\_new \gets []$; \\
    $HashCodes\_prev \gets \mathcal{I}.retrieve()$; (division excluded)\\
    
    \For{$v_i \in \mathcal{V}$}{
        $HashCodes.add({v_i})$; \\
    }

    $HashCodes \gets HashCodes\_new + HashCodes\_prev$; \\

    $W' \gets normalizeW(HashCodes)$; \\

    $HashCodes \gets divide(HashCodes, W')$; \\
    
    Hash Table: $T \gets construct(HashCodes)$; \\

    update $\mathcal{I}$: $T$, $W'$; \\
    
\end{algorithm}

\htitle{Update PQ Index.} \textbf{(Algorithm \ref{algorithm: dynamic pq update})} With those initial data, a product quantization (PQ) index has already been constructed. For newly added data, identifying the codebook of new points and iteratively updated the centroids, of affected cluster, at each subspace is sufficient for updating the product quantization index. In other words, such a simple updating method is efficient and achieves little degradation in terms of estimation. The index update rule applies for most of the datasets, however, for some datasets whose cardinality explodes for a tiny increase in distance threshold, the simple update rule aforementioned brings extra estimation error.

\begin{algorithm}[t]
    \caption{Dynamic Updates: PQ Index}
    \label{algorithm: dynamic pq update}
    \textbf{INPUT:} Existing PQ Index: $\mathcal{Q}$, New Points: $\mathcal{V} = \{v_i\}$;\\
    \textbf{OUTPUT:} Updated PQ Index: $\mathcal{Q}$; \\

    \For{$v \in \mathcal{V}$}{
        $v_1, v_2, ..., v_M$ = divide\_subspace(v);\\
        \For{$spID \in [1, 2,..., M]$}{
            $v_{spID}$ assigned to closest cluster;\\
        }
        Update $\mathcal{Q}$; \\
        1. In each subspace, update centroid of updated cluster; \\
    }
\end{algorithm}

\htitle{Update Neighbor Lookup Table.} \textbf{(Algorithm \ref{algorithm: update fast neighbor lookup})} Let's quickly review the functionality of neighbor look-up table $\mathcal{P}$. Technically, given a $\mathcal{B}_{cental}$ and a distance $k$, $\mathcal{P}$ enables the framework to efficiently retrieve all the hash buckets whose distance are at $k$ from $\mathcal{B}_{cental}$ in hamming space. Specifically, the process is just to compute the hamming distance between new hash codes and original hash codes, as well as the pair-wise distance among those new hash codes, which will be updated in the neighbor look-up table.

\begin{algorithm}[t]
    \caption{Dynamic Updates: Neighbor Lookup Table}
    \label{algorithm: update fast neighbor lookup}
    \textbf{INPUT:} Constructed table: $\mathcal{T}$, Array of existing hash codes: $\mathcal{C}$, Array of new hash codes: $\mathcal{C}_1$; \\
    \textbf{PARAMS:} $M$: distance greater than it will not be stored; \\
    \textbf{OUTPUT:} $\mathcal{T}'$: Updated Neighbor Lookup Table;\\

    $s_1, s_2 = \mathcal{C}.\text{size}(), \mathcal{C}_1.\text{size}()$; \\
    $s_{all} = s_1 + s_2$

    $\mathcal{C} \gets \mathcal{C} + \mathcal{C}_1$; \\

    $\mathcal{T}' \gets \mathcal{T}.\text{extend}(\mathcal{C}_1)$; \\
    
    \For{$0 \leq i <  s_1$}{
        \For{$s_1 \leq j < s_{all}$}{
            $d = dist_{hamming}(\mathcal{C}[i], \mathcal{C}[j])$; \\

            \If{$0 < d \leq M$}{
                $\mathcal{T}'[i].\text{update}\{(d, j)\}$; \\
                $\mathcal{T}'[j].\text{update}\{(d, i)\}$; \\
            }
        }
    }

    \For{$s_1 \leq i < s_{all}$}{
        \For{$s_1 \leq j < s_{all}$}{

            $d = dist_{hamming}(\mathcal{C}[i], \mathcal{C}[j])$; \\

            \If{$0 < d \leq M$}{

                $\mathcal{T}'[i].\text{update}\{(d, j)\}$; \\
            }
        }
    }
    \Return{$\mathcal{T}'$}; \\
\end{algorithm}

\section{Experiments}
\label{sec: evaluation}
We conduct extensive experiments on real world dataset for evaluation and analysis of our methods. We implement the Dynamic Prober~\footnote{https://github.com/OscarC9912/simQ\_hd\_card\_estimator} in C++ and compiled with g++ using \textsf{-Ofast} optimization. All experiments are conducted on a Ubuntu sever with Intel(R) Xeon(R) Gold $6248$ CPU ($160$ threads) and $1.5$ TB RAM. For model training requiring GPU, we use $\times 1$ NVIDIA A100 GPU (80GB, PCIE).

\subsection{Experimental Settings}

\htitle{Datasets.} We conduct the evaluation with $5$ real-world vector dataset, whose statistics are shown in Table \ref{tab:dataset_statistics}. All of these datasets are commonly used in the research of approximate nearest neighbor search\cite{10184454, DBLP:conf/icde/TianZZ22, DBLP:journals/pami/MalkovY20, DBLP:journals/pvldb/WangXY021} and cardinality estimation of similarity queries\cite{DBLP:conf/sigmod/Sun0021, 10.14778/3712221.3712224, DBLP:journals/pvldb/QinX18, DBLP:conf/sigmod/WangXQ0O00I21}. All of these datasets are acquired from the original source without extra processing.

\begin{table}[h]
\centering
\scalebox{1}{
\begin{tabular}{|c|c|c|c|c|c|}
\hline
\textbf{Dataset} & \textbf{Domain} & \textbf{\#Objects} & \textbf{Dimension} & \textbf{Test Size}\\
\hline
    SIFT & image & 1M & 128 & 1000 \\
\hline
    Glove & text & 2M & 300 & 2000 \\
\hline
    FastText & text & 1M & 300 & 1000 \\
\hline
    GIST & image & 1M & 960 & 1000 \\
\hline
    YouTube & video & 0.34M & 1770 & 340 \\
\hline
\end{tabular}}
\vspace{1mm}
\caption{Statistics of Datasets}
\label{tab:dataset_statistics}
\end{table}

\vspace{-4mm}

\htitle{Our Methods.} Dynamic Prober and Dynamic Probe-PQ are our methods involved in the evaluation, where dynamic prober-PQ is a version utilizing asymmetric distance computation in product quantization for point-wise distance computation.

\htitle{Competitors.} We compare the following approaches:

\begin{itemize}[left=0pt]
    \item \textbf{SimCard} \cite{DBLP:conf/sigmod/Sun0021}: a learning based cardinality estimation framework for similarity query featuring \textit{space segmentation} and \textit{query segmentation}. We adopt author's source code and follow the default settings for data generation and model training. 
    \item \textbf{MRCE} \cite{10.14778/3712221.3712224}: a learning based framework utilizing the knowledge of existing data called \textit{reference objects} for cardinality estimation. In order to comprehensively evaluate the performance of MRCE, we apply two settings by using $1\% \cdot |\mathcal{D}|$ and $10\% \cdot |\mathcal{D}|$ number of reference objects. We adopt author's source code and follow the default settings for data generation and model training.
    \item \textbf{Sampling $1\%$}: We uniformly sample $1\%$ of the data for estimation. Though simpleness, it is an effective approach and are widely used in the research of cardinality estimation problem, where \cite{DBLP:conf/sigmod/Sun0021, 10.14778/3712221.3712224} utilize uniform sampling of $1\%$ data as the competitor for sampling approach as well.
\end{itemize}

\htitle{Evaluation Metrics.} For accuracy, we follow the convention in recent research \cite{DBLP:conf/sigmod/Sun0021, 10.14778/3712221.3712224}, where we report the mean Q-Error as well as its distributions ($90\%$, $95\%$, $99\%$, and $100\%$ percentile). The Q-Error is defined as following: $$Q\text{-}error(\hat{c}, c) = \frac{max(\hat{c}, c)}{min(\hat{c}, c)},$$
where $c$ and $\hat{c}$ are real / estimated cardinality, respectively.

\htitle{Query Selection.}
In general, we generate the queries for evaluation following the conventions in recent works \cite{DBLP:conf/sigmod/Sun0021,10.14778/3712221.3712224, DBLP:conf/sigmod/WangXQ0O00I21}, where we partially adopted the code for data preprocessing from \cite{10.14778/3712221.3712224, DBLP:conf/sigmod/WangXQ0O00I21}. Specifically, for each dataset, we uniformly sample $K = min\{0.1\%\cdot|\mathcal{D}|, 1000\}$ of points as query vector, and for each of the query vector, we sample from a geometric sequence of $40$ values within the range of $[1, min(20000, 1\%\cdot|\mathcal{D}|)]$ as the ground truth cardinality, which bounds the smallest and largest cardinality. For each of the ground truth cardinality $c$, we find the minimum distance threshold $\tau$ that yields $c$ results, where $\tau$ is the distance threshold in the query.

\subsection{Estimation Accuracy}
Table \ref{table: estimation_accuracy} shows the Q-Error distribution of our methods as well as those competitors. The best results are highlighted in bold. Overall, we make the following observations: 1). Our method and reference CE$4$HD: MRCE are the most performant algorithms in terms of accuracy, and our dynamic prober achieves the best performance for the vast majority of evaluations. 2). Dynamic Prober (w./w.o PQ) outperforms all baseline methods (CE4HD: MRCE, SimCard, Sampling) in terms of mean and $90\%$ Q-error. 3). As for the $95\%$, $99\%$, and maximum Q-error, the dynamic prober outperforms other method except for two datasets: SIFT and FastText, where our Q-error is slightly higher than the optimal. 4). The product quantization accelerated dynamic prober achieves similar or even better accuracy than the method without acceleration across most of the dataset. However, for GloVe and FastText, we can observe some degradation in terms of accuracy when distance is estimated with product quantization. The degradation in performance can be attributed to property of dataset, where the dataset suffers from a sudden and significant change in distance distribution, as revealed by previous work\cite{10.14778/3712221.3712224}; and the estimation will bring some more noticeable degradation in Q-error for a slight error in distance estimation. 5). SimCard and Sampling $1\%$ achieves the worst performance for all the time. Despite sampling $1\%$ of the dataset introduced significant error, its performance is relatively more stable than SimCard.

\begin{table}[t]
\centering
\label{tab:performance}
\scalebox{0.83}{
\begin{tabular}{|c|c|c|c|c|c|c|}
\hline
\textbf{Dataset} & \textbf{Method} & \textbf{Mean} & \textbf{90th} & \textbf{95th} & \textbf{99th} & \textbf{Max} \\
\hline
\multirow{6}{*}{SIFT} & CE4HD: MRCE 1\% & 2.11 & 3.33 & 4.39 & 8.25 & 71 \\
 & CE4HD: MRCE 10\% & 1.59 & 2.34 & \textbf{2.95} & \textbf{4.56} & \textbf{13.22} \\
 & SimCard: GL+ & 6.2 & 11.24 & 18 & 56.78 & 537.3 \\
 & Sampling 1\% & 12.3 & 35 & 66 & 158 & 585 \\
 & Dynamic Prober & \textbf{1.56} & \textbf{2.25} & 3 & 6 & 21.5 \\
 & Dynamic Prober-PQ & 1.69 & 2.56 & 3.6 & 7.5 & 51 \\
\hline
\multirow{6}{*}{Glove} & CE4HD: MRCE 1\% & 4.92 & 9.26 & 12.31 & 34.33 & 742.67 \\
 & CE4HD: MRCE 10\% & 4.45 & 9.05 & 11.84 & 27.67 & 587.67 \\
 & SimCard: GL+ & 242.9 & 417.1 & 1040 & 5920 & 899691 \\
 & Sampling 1\% & 11.2 & 27 & 54 & 138 & 565 \\
 & Dynamic Prober & \textbf{2.9} & \textbf{4.78} & \textbf{7} & \textbf{19.4} & \textbf{223} \\
 & Dynamic Prober-PQ & 4.19 & 7.37 & 11.77 & 27 & 282.5 \\
\hline
\multirow{6}{*}{FastText} & CE4HD: MRCE & 2.31 & 4.18 & 5.18 & 8.14 & 32.41 \\
 & CE4HD: MRCE 10\% & 2.06 & 3.46 & \textbf{4.33} & \textbf{7.25} & \textbf{40.42} \\
 & SimCard: GL+ & 50.6 &	86.8 & 197 & 728 & 10000 \\
 & Sampling 1\% & 12.87 & 35 & 66 & 158 & 585 \\
 & Dynamic Prober & \textbf{1.99} & \textbf{3} & 5 & 11.5 & 235.5 \\
 & Dynamic Prober-PQ & 2.59 & 4.82 & 7.5 & 15 & 98.5\\
\hline
\multirow{6}{*}{GIST} & CE4HD: MRCE 1\% & 10.62 & 9.64 & 18.64 & 158.87 & 2200.85 \\
 & CE4HD: MRCE 10\% & 14.42 & 9.59 & 18.57 & 244.36 & 2699.32 \\
 & SimCard: GL+ & 17.5 & 37.8 & 72.4 & 213.5 & 2471.8 \\
 & Sampling 1\% & 12.43 & 35 & 66 & 158 & 728 \\
 & Dynamic Prober & 4.4 & 8.2 & 14 & 33 & \textbf{378} \\
 & Dynamic Prober-PQ & \textbf{4} & \textbf{7.53} & \textbf{10.71} & \textbf{20.8} & 728 \\
\hline
\multirow{6}{*}{YouTube} & CE4HD: MRCE 1\% & 3.82 & 7.67 & 9.5 & 15.72 & 70.62 \\
 & CE4HD: MRCE 10\% & 3.89 & 8.16 & 10.39 & 17.11 & 49.33 \\
 & SimCard: GL+ & 7.72 & 15.69 & 30.25 & 92 & 423 \\
 & Sampling 1\% & 13.3 & 36 & 63 & 163 & 512 \\
 & Dynamic Prober & 2.56 & 4.31 & 6 & 12.5 & \textbf{26} \\
 & Dynamic Prober-PQ & \textbf{2.08} & \textbf{3.5} & \textbf{5} & \textbf{10.2} & \textbf{26} \\
\hline
\end{tabular}
}
\caption{Performance: Q-Error Distribution}
\label{table: estimation_accuracy}
\end{table}

\subsection{Efficiency: Offline Data Preparation and Model Construction}

\begin{figure}[t]
    \centering
    \includegraphics[width=\linewidth]{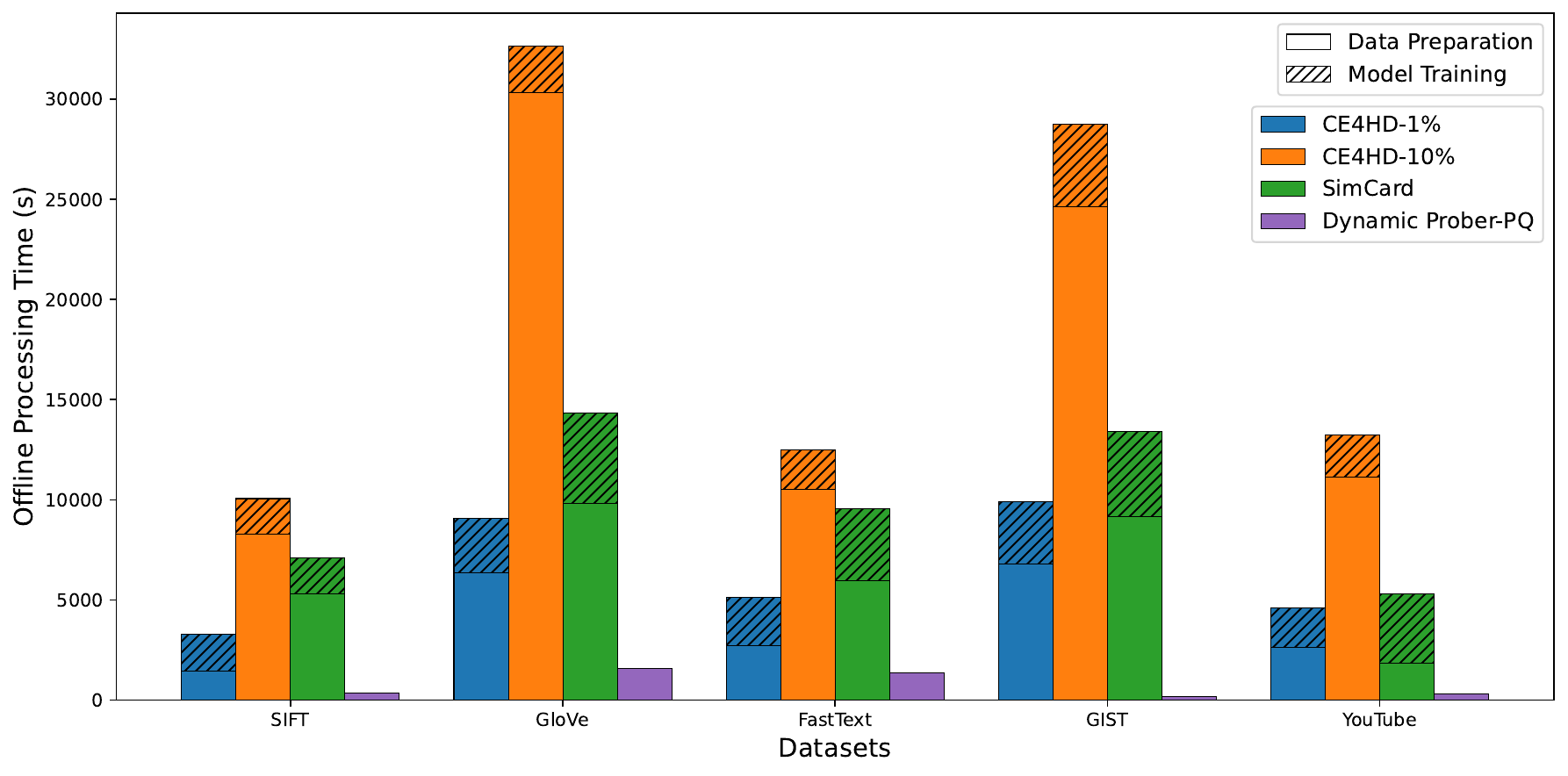}
    \vspace{-2em}
    \caption{Efficiency: Time of Offline Estimator Construction (include all phase of construction of each methods)}
    \label{fig: compare offline latency}
\end{figure}

Learning-based methods require substantial time both for constructing training datasets and for the training phase itself, even when leveraging modern GPUs. In short, building a learned model for prediction is highly time-consuming. By contrast, our estimator—based on locality sensitive hashing (LSH)—can be constructed with significantly less overhead. We quantitatively demonstrate this advantage in Figure \ref{fig: compare offline latency}, which clearly shows that the offline processing latency of our method is much lower than that of all other approaches, making it the most efficient in terms of offline construction.

\begin{figure}[t]
    \centering
    \includegraphics[width=\linewidth]{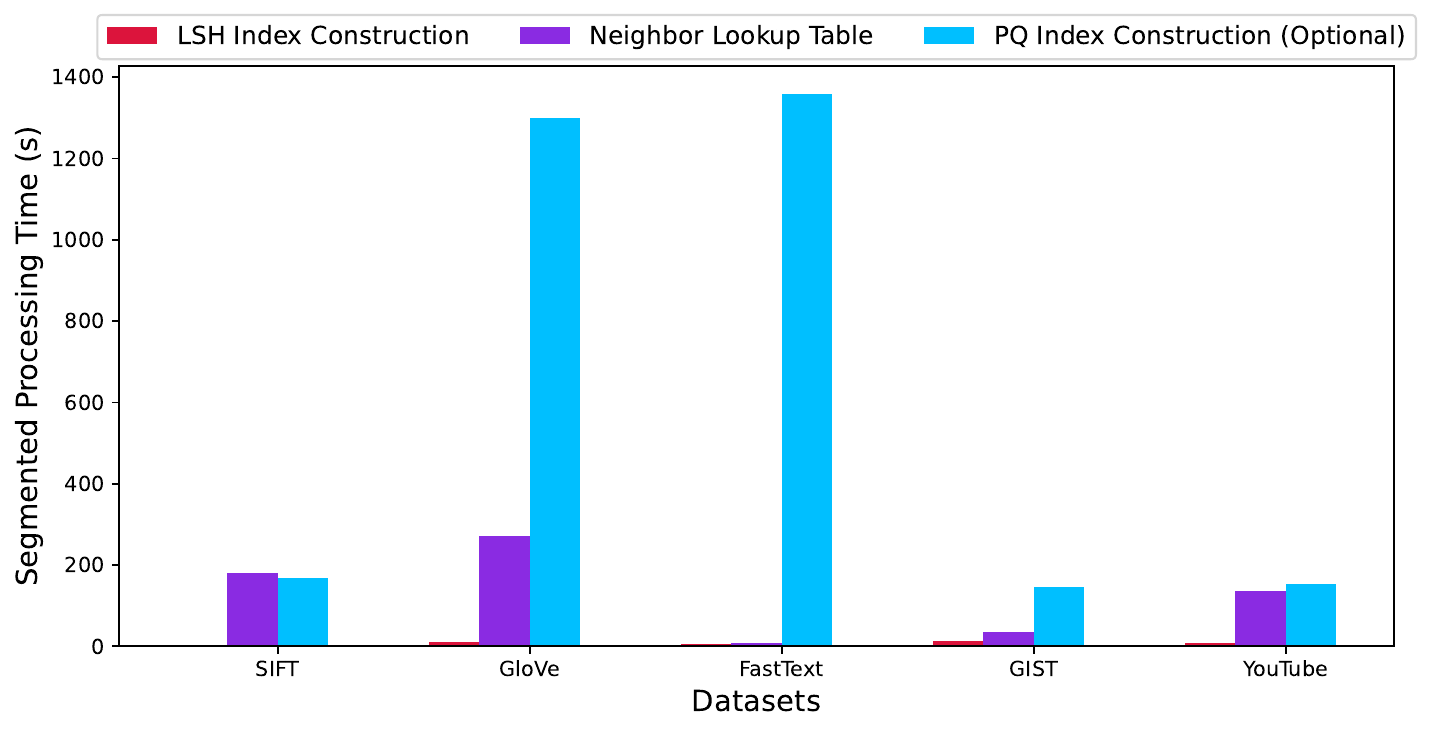}
    \vspace{-2em}
    \caption{Dynamic Prober: Break down time of offline construction, containing all $3$ phases}
    \label{fig: offline efficiency segmented}
\end{figure}

For our method, the offline construction latency stems from three components: (i) building the LSH index, (ii) generating the neighbor look-up table, and (iii) optionally constructing a product quantization (PQ) index for asymmetric distance estimation. Importantly, no additional overhead is incurred in processing the dataset itself. As shown in Figure \ref{fig: offline efficiency segmented}, the time required for both LSH index construction and neighbor look-up table generation is indeed small, while the optional PQ index constitutes the dominant portion of the offline construction latency for our method.

\subsection{Efficiency: Online Estimation}

Table \ref{table: online efficiency} reports the latency of online estimation for our method and the baselines. We highlight three key observations. 1). Among learning-based approaches, CE4HD: MRCE achieves the lowest latency by leveraging both the efficiency of neural networks and reference objects. 2). Among non-learning-based approaches, Dynamic Prober demonstrates competitive latency, performing comparably to sampling-based methods while offering superior efficiency in high-dimensional datasets such as GIST ($960$D) and YouTube ($1770$D). 3). The incorporation of asymmetric distance computation further enhances efficiency of our method, contributing additional acceleration during online estimation.

\begin{table}[t]
\centering
\scalebox{0.88}{
\begin{tabular}{|l|c|c|c|c|c|c|}
\hline
\textbf{Method} & \textbf{SIFT1M} & \textbf{Glove} & \textbf{FastText} & \textbf{GIST} & \textbf{YouTube}\\
\hline
MRCE $1\%$ & 6.4 & 4.2 & 4.2 & 3.9 & 4.8 \\
\hline
SimCard & 49.1 & 53.4 & 46.8 & 66.4 & 72.3 \\
\hline
Sampling $1\%$ & 89 & 217 & 113 & 206 & 391\\
\hline
DynamProber & 59 & 235 & 138 & 51 & 77\\
\hline
DynamProber-PQ & 46 & 213 & 121 & 33 & 58 \\
\hline
\end{tabular}
}
\caption{Efficiency: Time of Online Estimation (ms)}
\label{table: online efficiency}
\end{table}

\subsection{Efficiency: Asymmetric Distance Estimation}
\begin{figure}[t]
    \centering
    \includegraphics[width=\linewidth]{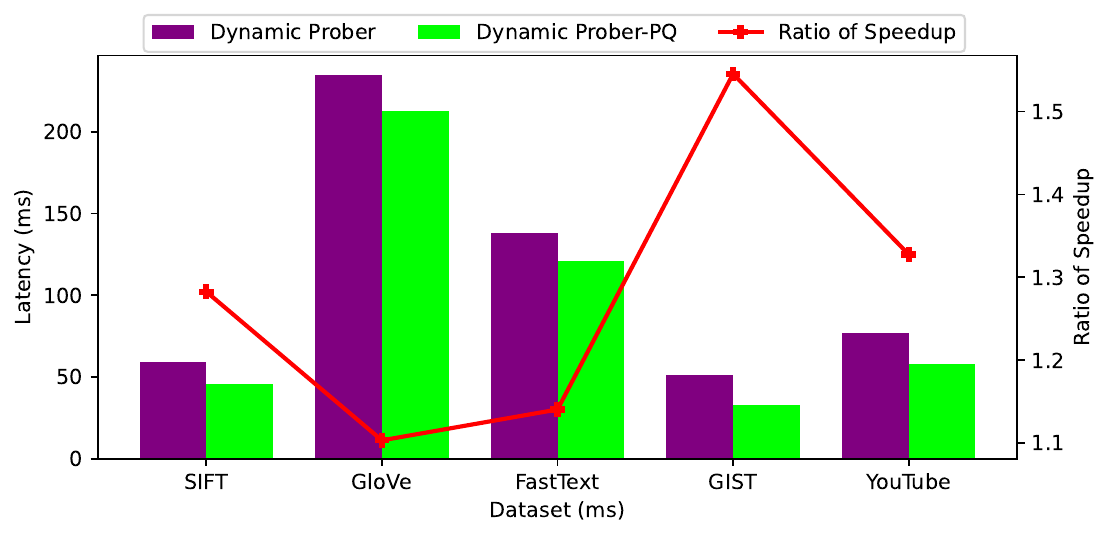}
    \vspace{-2em}
    \caption{Dynamic Prober V.S. Dynamic Prober-PQ: Speedup}
    \label{fig: ablation pq}
\end{figure}

In this section, we present an ablation study evaluating the benefits of asymmetric distance computation with a product quantization index, with results shown in Figure \ref{fig: ablation pq}. As illustrated, asymmetric distance computation improves computational efficiency, achieving a speedup of approximately $\times 1.6$. Furthermore, the benefits of distance estimation increase with the dimensionality of the dataset, indicating that this approach is particularly advantageous for high-dimensional datasets.

\subsection{Parameter Study: Error Tolerance Value $\epsilon$}

\begin{figure}[t]
    \centering
    \includegraphics[width=\linewidth]{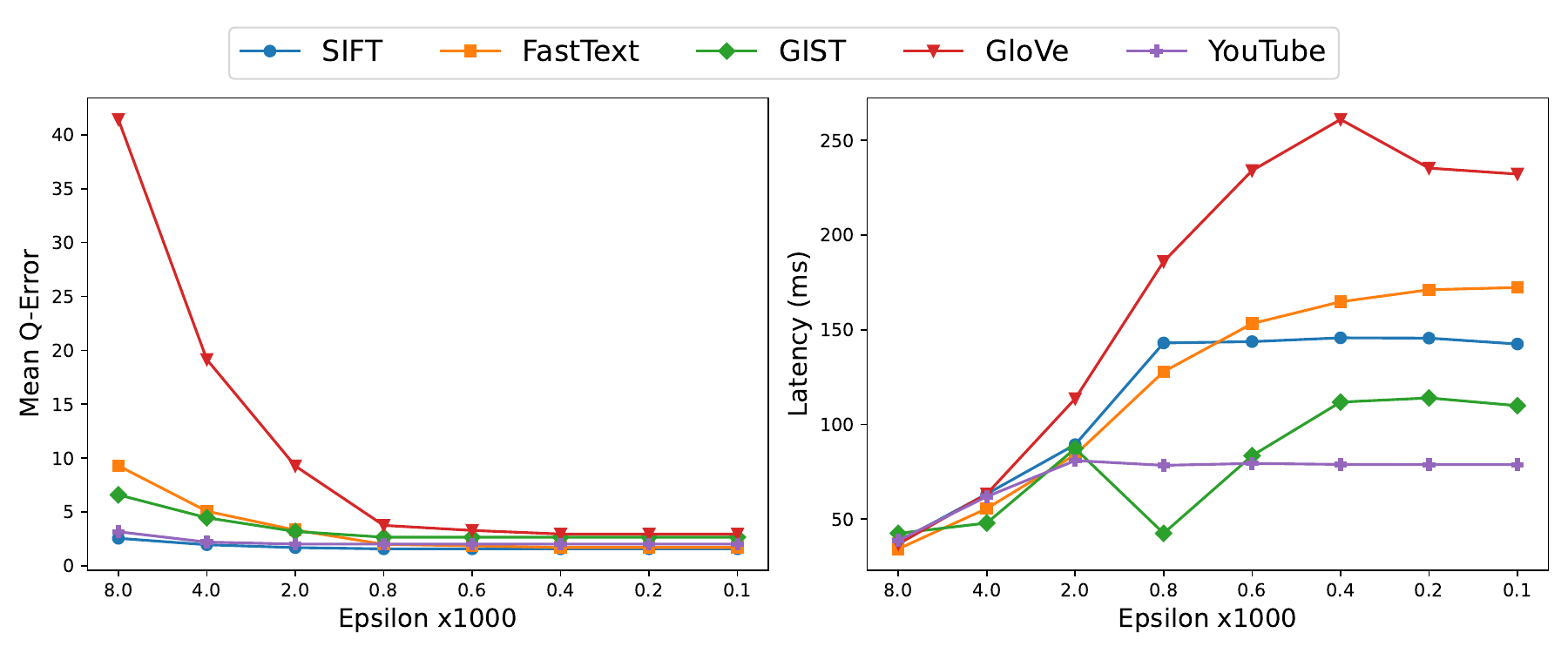}
    \vspace{-2em}
    \caption{Parameter Study - $\epsilon$: Accuracy and Efficiency}
    \label{fig: parameter study espsilon}
\end{figure}

In this section, we investigate into how the hyperparameter $\epsilon$ influences the trade off between efficiency and accuracy, whose result is demonstrated with Figure \ref{fig: parameter study espsilon}. Before getting into the experimental result, let's briefly review the functionality of the $\epsilon$. Theoretically, the parameter $\epsilon$ reflects the estimation error allowed. More practically, in our neighboring-based probing scheme, the $\epsilon$ controls the following features: 1). necessity of estimating with larger samples in progressive sampling; 2). necessity of exploring neighboring buckets that are most distant.

In Figure \ref{fig: parameter study espsilon}, a larger value of $\epsilon$ demonstrates better efficiency, but the error can be more significant, as the estimator neither samples sufficient amount of points nor exploring neighbors that are distant enough. On the contrary, a smaller $\epsilon$ gives better accuracy but higher latency. However, it is worthwhile noticing that there is no need for the $\epsilon$ to be arbitrarily small. As we can observe from Figure \ref{fig: parameter study espsilon}, the mean Q-error no longer improves after some turning point, which varies among datasets and can be tuned for each dataset.

\begin{figure}[t]
    \centering
    \includegraphics[width=1\linewidth]{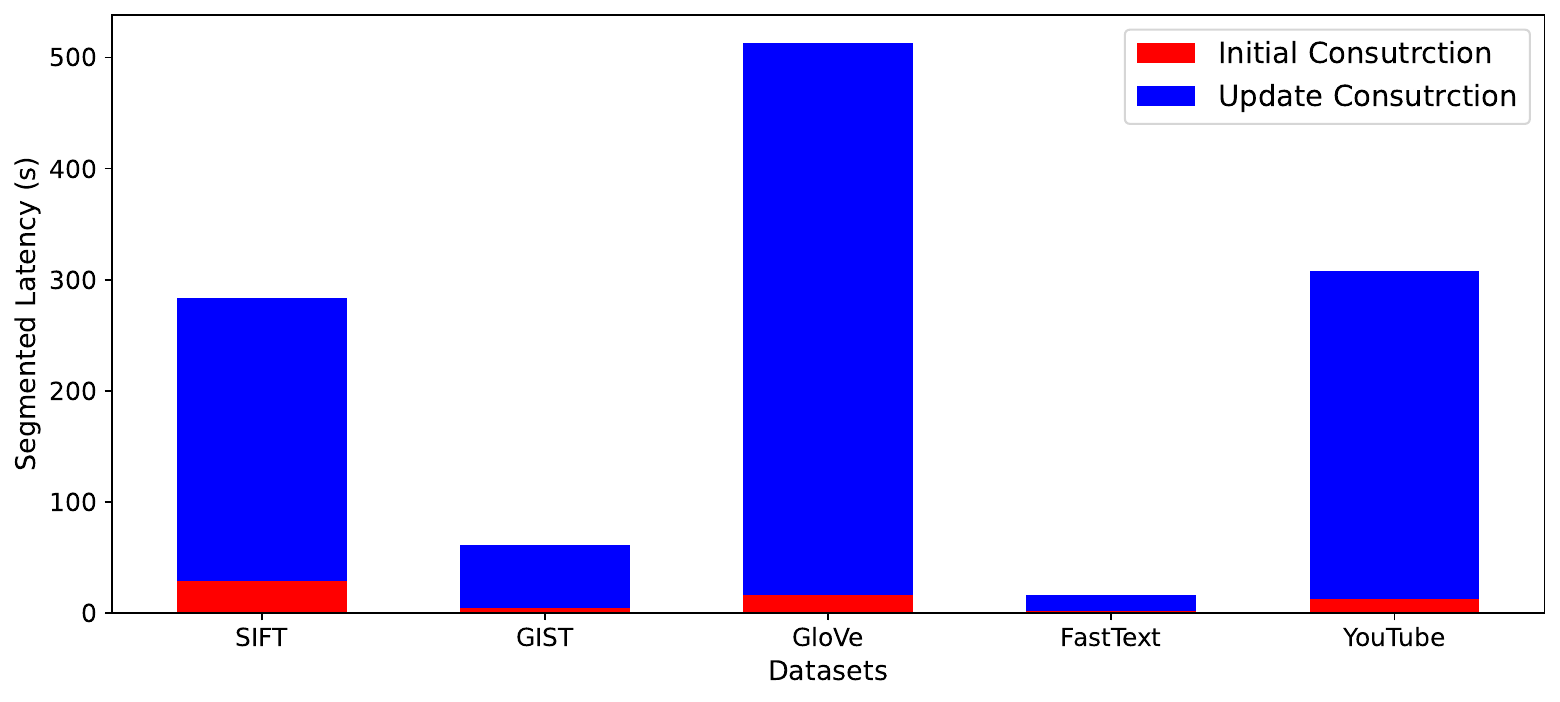}
    \caption{Dynamic Prober: Dynamic Update Efficiency ($10\%$ data for initial framework construction and $90\%$ data for framework update)}
    \label{fig:update efficiency}
\end{figure}

\begin{figure}[t]
    \centering
    \includegraphics[width=1\linewidth]{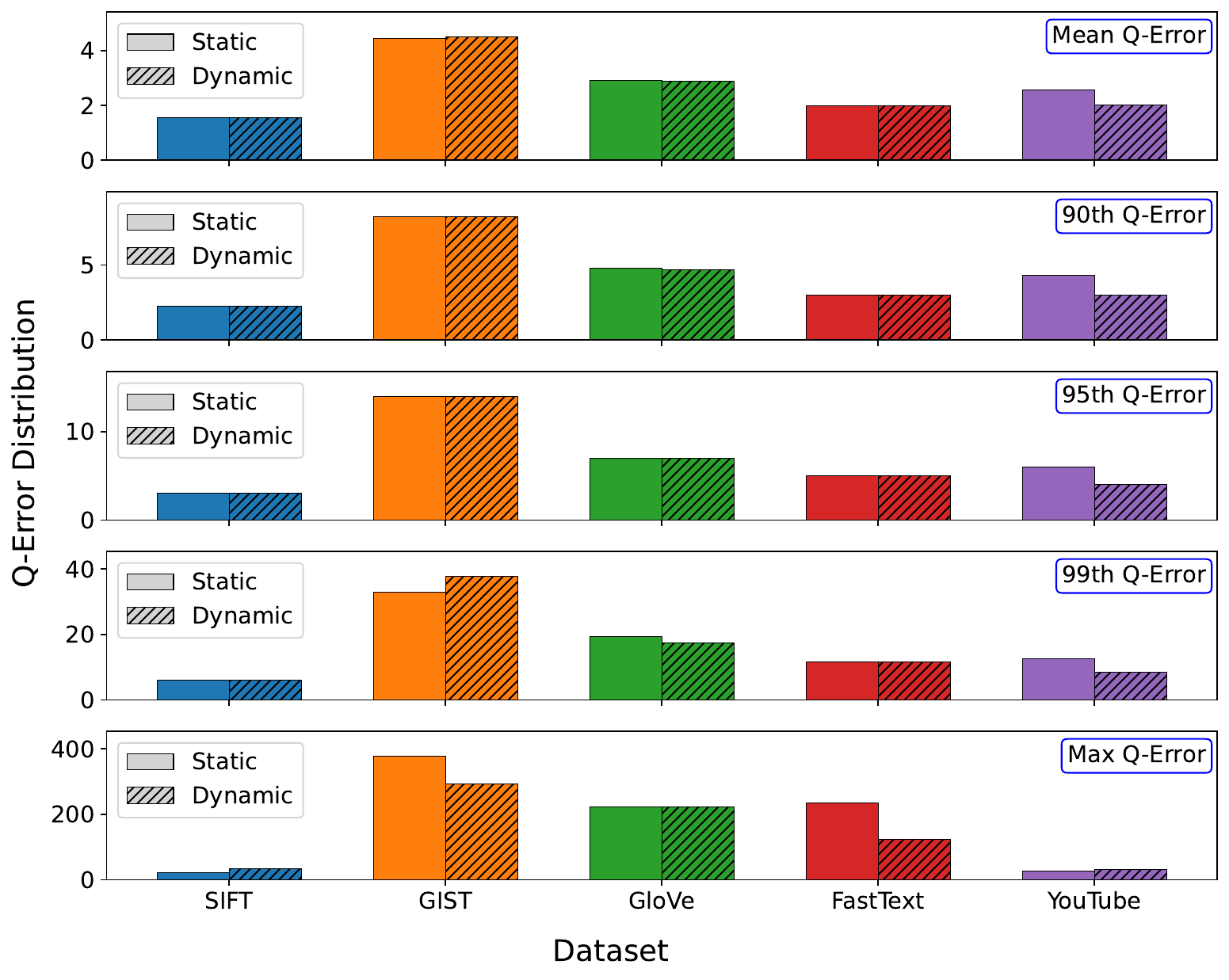}
    \caption{Dynamic Prober: Dynamic Update Accuracy (The shadowed bars refers to the case of dynamic data updates, where $10\%$ data is used for initial framework construction and $90\%$ data is used for framework updates.)}
    \label{fig:update accuracy}
\end{figure}

\begin{table}[t]
\centering
\label{tab:performance}
\scalebox{0.85}{
\begin{tabular}{|c|c|c|c|c|c|c|}
\hline
\textbf{Dataset} & \textbf{Method} & \textbf{Mean} & \textbf{90th} & \textbf{95th} & \textbf{99th} & \textbf{Max} \\
\hline
\multirow{3}{*}{SIFT} & CE4HD: MRCE 1\% & 2156 & 8656 & 13398 & 16668 & 16668 \\
 & CE4HD: MRCE 10\% & 12 & 22 & 33 & 75 & 784 \\
 & Dynamic Prober & \textbf{1.56} & \textbf{2.25} & \textbf{3} & \textbf{6} & \textbf{33}\\
\hline
\multirow{3}{*}{Glove} & CE4HD: MRCE 1\% & 3800 & 12479 & 19971 & 31957 & 31957 \\
 & CE4HD: MRCE 10\% & 3801 & 12479 & 19971 & 31957 & 31957 \\
 & Dynamic Prober & \textbf{2.88} & \textbf{4.67} & \textbf{7} & \textbf{17.4} & \textbf{223}\\
\hline
\multirow{3}{*}{FastText} & CE4HD: MRCE 1\% & 48 & 156 & 237 & 361 & 515 \\
 & CE4HD: MRCE 10\% & 27 & 82 & 115 & 172 & 482 \\
 & Dynamic Prober & \textbf{1.99} & \textbf{3} & \textbf{5} & \textbf{11.5} & \textbf{122.5} \\
\hline
\multirow{3}{*}{GIST} & CE4HD: MRCE 1\% & 2055 & 6958 & 10769 & 16668 & 16668 \\
 & CE4HD: MRCE 10\% & 2055 & 6958 & 10770 & 16668 & 16668 \\
 & Dynamic Prober & \textbf{4.51} & \textbf{8.22} & \textbf{14} & \textbf{37.8} & \textbf{292.5} \\
\hline
\multirow{3}{*}{YouTube} & CE4HD: MRCE 1\% & 840 & 3251 & 4766 & 5769 & 5769 \\
 & CE4HD: MRCE 10\% & 840 & 3251 & 4766 & 5769 & 5769 \\
 & Dynamic Prober & \textbf{2.0} & \textbf{3} & \textbf{4} & \textbf{8.5} & \textbf{31.5} \\
\hline
\end{tabular}
}
\caption{Competitors: Dynamic Update Accuracy ($10\%$ data for initial framework construction and $90\%$ data for update)}
\label{table: dynamic competitor accuracy}
\end{table}

\subsection{Evaluation: Data Updates}

In this section, we evaluate the performance of data update scheme.

\htitle{Generate Dynamic Dataset:} In this experiment, different from updating hundreds of points in a million level dataset, we conduct the evaluation under a large-scale data update. Specifically, the dynamic dataset is constructed entirely based on previous static datasets $\mathcal{D}$, where we uniformly sample $10\%$ of $\mathcal{D}$ as the initial dataset, and use the rest of $90\%$ of dataset for updating. In addition, queries are completely the same as queries in the static case.

\htitle{Select Competitors:} In the evaluation of data updates, we select the MRCE$1\%$ and MRCE$10\%$, which is most competitive framework in static scenario, as our major competitors. As a learning-based cardinality estimator, supporting a dynamic data update scheme can be mainly achieved by two approach: 1). directly utilize the original estimation model for updated data; 2). re-generate the training data (partial or complete) and re-train the model. In the evaluation for the offline construction time, we observed that constructing a new model can be time-consuming, and it is easy to infer that the reconstruction of a updated model is also time-consuming, which is infeasible for practical purposes. To tackle this, in this evaluation, we directly use original models to estimate on the updated dataset. 

\htitle{Evaluation of Efficiency}: For dynamic prober, we use the algorithm update algorithms discussed in Section \ref{sec: dynamic data} to support data updating scheme, where we evaluate the accuracy and efficiency of data update. Specifically, Figure \ref{fig:update efficiency} demonstrates the efficiency of updating of our method, where we can make the following observations: 1). the initial construction of our framework takes very small amount of time, as the initial data is merely $10\%$ of the dataset; and the time to update the constructed framework on $90\%$ of the dataset takes higher but acceptable amount of time. In short, our approach takes a reasonable amount of time for updating large scale dataset based on the initial framework constructed with initial dataset.

\htitle{Evaluation of Accuracy}: We conduct the evaluation for accuracy by comparing our dynamic update scheme with the static one, where we construct the framework with all the dataset, as well as our competitors aforementioned. Figure \ref{fig:update accuracy} demonstrates the comparison of Q-error distribution between static dataset and two stages update scheme, where we can observe that there is little degradation in accuracy with our updating scheme. What's more, we can also observe some slight improvement in Q-error with the two stages construction, potentially due to better index construction. In short, our data updating algorithm will not incur noticeable degradation in the accuracy of estimation.

Table \ref{table: dynamic competitor accuracy} demonstrates how our competitors performs with a large-scale dynamic data update scheme, whose prediction is estimated with the original model constructed by initial dataset. We can make the following observations: 1). overall, the Q-error is significantly higher than the static case, indicating that utilizing original model cannot provide accurate estimation under a large-scale data updating scheme; 2). utilizing a larger amount of reference objects (from $1\%$ to $10\%$) cannot always guarantee a better performance for all the datasets, as the extra data included might contribute little to the training data and thus model performance, which also demonstrates that the performance of learning based model relies heavily on the quality of training data.

\section{Conclusion}
\label{sec: conclusion}
In this paper, we study the problem of cardinality estimation for similarity search in high dimensional space. Leveraging the locality-sensitive hashing index in approximate nearest neighbor search, we propose a new concept dynamic prober to adaptively probe the neighboring buckets to estimate the cardinality of similarity queries. What's more, we further optimize the efficiency of algorithm by applying progressive sampling and the asymmetric distance computation in product quantization. We conduct extensive experiments demonstrating the superiority in performance of our approach. To conclude, we propose three potential future directions of our work: (1). explore a more unified non-learning framework supporting the estimation in different distance spaces; (2). apply our method to support optimization tasks in downstream tasks related to semantic operators or vector database; (3). explore other potential training-less methods to estimate the cardinality of similarity queries. 

\bibliographystyle{ACM-Reference-Format}
\bibliography{myRef}

\newpage

\onecolumn
\section{Appendix}
\vspace{2mm}
\subsection{Definition: Chernoff Bound}

\vspace{1em}

\noindent{\textbf{Consider tossing an unfair coin for $n$ times, we have:}}

\begin{itemize}
    \item Observation: $X_i$
        \begin{itemize}
            \item The outcome of the $i^{\text{th}}$ toss
            \item 1 for Heads, 0 for Tails
        \end{itemize}
    \item True Probability: $p$
        \begin{itemize}
            \item The actual probability of obtaining Heads
        \end{itemize}
    \item Error parameter: $\epsilon$
        \begin{itemize}
            \item The margin of error for our estimation.
        \end{itemize}
\end{itemize}

\vspace{2em}

\textbf{\textit{The Chernoff Bounds are represented as following:}}

\[
Pr\left(\frac{1}{n}\sum X_i < p - \epsilon\right) \leq exp\left(-\frac{\epsilon^2 n}{2p}\right)
\]

\vspace{1em}

\[
Pr\left(\frac{1}{n}\sum X_i > p + \epsilon\right) \leq exp\left(-\frac{\epsilon^2 n}{2p + \frac{2\epsilon}{3}}\right)
\]

\newpage

\vspace{2mm}

\subsection{Proof: $\Pr(p > \mu_{upper}) \geq 1 - \delta$}

\vspace{1mm}

\textbf{Notation:}
\begin{itemize}
    \item $1-\delta$: probability of bounding the upper bound of $p$ with $\mu_{upper}$
    \item $a = \ln\left(\frac{1}{\delta}\right)$
\end{itemize}

\noindent\textbf{Want to Show:}
\[
\Pr(P < \mu_{\text{upper}}) \geq 1 - \delta
\]

\noindent\textbf{Prove the other side:}
\[
\Pr(P > \mu_{\text{upper}}) \leq \delta
\]

\begin{align}
&\Pr(p > \mu_{\text{upper}}) = \Pr\left(p > \left(\sqrt{\hat{p} + \frac{a}{2w}} + \sqrt{\frac{a}{2w}}\right)^{\,2}\right) \tag{1}\\[0.8em]
&= \Pr\left[p > \hat{p} + \frac{a}{2w} + \frac{a}{2w} + 2\sqrt{\frac{a}{2w}} \cdot \sqrt{\hat{p} + \frac{a}{2w}}\right] \tag{2}\\[0.8em]
&\leq \Pr\left[p - 2\sqrt{\frac{a}{2w}}\sqrt{\hat{p} + \frac{a}{2w}} + \frac{a}{2w} > \hat{p} + \frac{2a}{2w}\right] \tag{3}\\[0.8em]
&\leq \Pr\left[p - 2\sqrt{\frac{a}{2w}}\sqrt{p} + \frac{a}{2w} > \hat{p} + \frac{2a}{2w}\right] \tag{4}\\[0.8em]
&\leq \Pr\left[\left(\sqrt{p} - \sqrt{\frac{a}{2w}}\right)^2 > \hat{p} + \frac{2a}{2w}\right] \tag{5}\\[0.8em]
&= \Pr\left[p - 2\sqrt{p}\sqrt{\frac{a}{2w}} + \frac{a}{2w} > \hat{p} + \frac{2a}{2w}\right] \tag{6}\\[0.8em]
&\leq \Pr\left[p - 2\sqrt{\frac{ap}{2w}} > \hat{p} + \frac{a}{2w}\right] \tag{7}\\[0.8em]
&= \Pr\left[-2\sqrt{\frac{aP}{2w}} - \frac{a}{2w} > \hat{p} - p\right] \tag{8}\\[0.8em]
&\leq \Pr\left[-2\sqrt{\frac{ap}{2w}} > \hat{p} - p\right] \tag{by Chernoff Bound}\\[0.8em]
& \leq \exp\left(-\frac{4 \cdot \frac{ap}{2w} \cdot w}{2 \cdot p}\right) = \exp(a) = \exp(\ln(\frac{1}{\delta})) = \delta \tag{10}
\end{align}

\noindent Therefore, we have shown that:
\[
\Pr(p > \mu_{\text{upper}}) \leq \delta
\]

\noindent which is equivalent to:
\[
\Pr(p < \mu_{\text{upper}}) \geq 1 - \delta
\]

\vspace{2mm}

\textbf{\textit{To conclude, according to what we have proved, we are $1-\delta$ confident that the upper bound of real selectivity $p$ can be bounded by $\mu_{upper}$.}}

\end{document}